\documentclass[twocolumn,showpacs,nofootinbib,preprintnumbers,amsmath,amssymb,showkeys,superscriptaddress]{revtex4-1}

\usepackage[pdftex]{graphicx}  
\usepackage{color}
\usepackage{slashed}
\usepackage[normalem]{ulem}
\usepackage{braket}
\usepackage{dsfont}

\newcommand{\cb}{\overline c}

\newcommand\eqn[1]{(\ref{#1})}      
\newcommand\Eqn[1]{Eq.~(\ref{#1})}  
\newcommand\Fig[1]{Fig.~\ref{#1}}  

\newcommand \beq{\begin{eqnarray}}
\newcommand \eeq{\end{eqnarray}}

\begin{document}

\title{Small parameters in infrared QCD: The pion decay constant}

\author{Marcela Pel\'aez\vspace{.4cm}}%
\affiliation{%
Instituto de F\'{\i}sica, Facultad de Ingenier\'{\i}a, Universidad de
la Rep\'ublica, J. H. y Reissig 565, 11000 Montevideo, Uruguay.
\vspace{.1cm}}%
\author{Urko Reinosa}%
\affiliation{%
Centre de Physique Th\'eorique, CNRS, Ecole polytechnique, IP Paris, F-91128 Palaiseau, France
\vspace{.1cm}}%
\author{Julien Serreau}%
\affiliation{ Universit\'e Paris Cit\'e, CNRS, Astroparticule et Cosmologie, F-75013 Paris, France.\vspace{.1cm}}
\author{Nicol\'as Wschebor}%
\affiliation{%
 Instituto de F\'{\i}sica, Facultad de Ingenier\'{\i}a, Universidad de
 la Rep\'ublica, J. H. y Reissig 565, 11000 Montevideo, Uruguay.
\vspace{.1cm}}%

\date{\today}

\begin{abstract}
We continue our investigation of the QCD dynamics in terms of the Curci-Ferrari effective Lagrangian, a deformation of the  Faddeev-Popov one in the Landau gauge with a tree-level gluon mass term. In a previous work we have studied the dynamics of chiral symmetry breaking at the level of the quark propagator and, in particular, the dynamical generation of a constituent quark mass. In the present article, we study the associated Goldstone mode, the pion, and we compute the pion decay constant in the chiral limit. Our approach exploits the fact that the coupling (defined in the Taylor scheme) in the pure gauge sector is perturbative, as observed in lattice simulations which, together with a $1/N_c$-expansion, allows for a systematic, controllable approximation scheme in the low energy regime of QCD. At leading order, this leads to the well-known rainbow-ladder resummation. We study the region of parameter space of the model that gives physical values of the pion decay constant. This allows one to constrain the gluon mass parameter as a function of the coupling using a physically measured quantity.   
\end{abstract}

\maketitle

\section{Introduction}

The most prominent aspects of the QCD dynamics at large distances, namely confinement and dynamical chiral symmetry breaking, are of intrinsic nonperturbative nature in terms of the elementary (quark and gluon) degrees of freedom of the theory. This common wisdom has two aspects. The first, rather trivial one is simply that describing phenomena such as bound states of quarks and gluons (as required by confinement) or dynamical quark mass generation (as implied by chiral symmetry breaking) require resumming diagrams at infinite loop orders.  The second, deeper aspect follows from the fact that the standard perturbative approach, based on the Faddeev-Popov (FP) Lagrangian, predicts a Landau pole, where the running coupling constant diverges, and is thus not applicable in the infrared regime.  Even though the first problem can be overcome, at least in principle, by standard resummation techniques, as done, e.g., to describe QED bound states, the second problem kills this hope because of the lack of a proper expansion scheme to select the diagrams to be resummed.

The above, apparently hopeless description, however, suffers from a serious loophole. On the formal level, first, the FP approach to gauge theories is known to be plagued by the issue of Gribov ambiguities \cite{Gribov:1977wm,Zwanziger:1981kg}, which inherently limits its validity to, at best, the deep ultraviolet (UV) regime. In fact, no nonperturbative version of the FP gauge-fixed Lagrangian (say, in the Landau gauge) or of any BRST-invariant Lagrangian has been constructed so far \cite{Neuberger:1986vv,Neuberger:1986xz}. Moreover, on a practical level, actual lattice calculations of gauge-dependent quantities in the (lattice) Landau gauge\footnote{Existing lattice gauge fixing procedures involve an extra ingredient than the sole (e.g., Landau) gauge fixing condition in order to solve the Gribov problem. For instance, one explicitly selects one Gribov copy or one averages over a subset of copies, etc. It is not known, however, how to formulate such procedures by means of a local, renormalizable, gauge-fixed action.} have revealed stringent features of the infrared QCD dynamics \cite{Mandula:1987rh,Bonnet:2000kw,Bonnet:2001uh,Cucchieri:2007rg,Bogolubsky:2009dc,Bornyakov:2009ug,Iritani:2009mp,Boucaud:2011ug,Maas:2011se,Oliveira:2012eh,Bowman:2004jm,Bowman:2005vx,Silva:2010vx}. In the pure gauge sector, one observes, first, that the gluon propagator saturates at vanishing (Euclidean) momentum, signaling the dynamical generation of a nonzero screening mass (whereas the ghost propagator remains massless) and, second, that the coupling constant is finite in the infrared, showing no sign of a Landau pole. In fact, the (Taylor) coupling in the pure gauge sector remains moderate at infrared momenta and even vanishes in the deep infrared. This strongly advocates for the possibility of a modified perturbative approach to infrared QCD dynamics.

As a completely justified gauge-fixed Lagrangian in the continuum is still lacking,\footnote{Quantization procedures which aim at solving the Gribov issue of the FP approach have been proposed \cite{Zwanziger:1989mf,Serreau:2012cg,Serreau:2013ila,Dudal:2007cw}, although none of them is completely satisfactory so far.} one can resort to model Lagrangians motivated by phenomenological (lattice) observations. 
The simplest such proposal \cite{Tissier:2010ts,Tissier:2011ey} consists in adding a bare gluon mass term to the FP Lagrangian (in the Landau gauge), which is a particular case of the class of Curci-Ferrari (CF) Lagrangians \cite{Curci:1975cw,Curci:1976bt}. Such a soft deformation of the FP theory remains perturbatively renormalizable and does not modify the well-tested ultraviolet regime of the theory. Most importantly, the model possesses infrared safe renormalization group trajectories, with no Landau pole \cite{Tissier:2011ey,Weber:2011nw,Reinosa:2017qtf,DallOlio:2020xpu}, allowing for a well-defined perturbative expansion down to arbitrary infrared scales. A large body of work in the past decade has put this modified perturbative approach to test and has demonstrated that it efficiently captures various aspects of the infrared dynamics of both Yang-Mills theories and QCD-like theories with heavy quarks \cite{Tissier:2010ts,Tissier:2011ey,Pelaez:2013cpa,Pelaez:2014mxa,Reinosa:2014ooa,Reinosa:2014zta,Pelaez:2015tba,Reinosa:2015gxn,Gracey:2019xom,Barrios:2020ubx,Pelaez:2021tpq,Figueroa:2021sjm,Barrios:2021cks,vanEgmond:2021jyx,Barrios:2022hzr}. One- and, in some cases, two-loop calculations of numerous infrared sensitive quantities (two- and three-point functions, phase diagram at nonzero temperature and densities, etc.)  compare very well with actual lattice calculations. The CF model also leads to interesting neutron star phenomenology \cite{Song:2019qoh,Suenaga:2019jjv,Kojo:2021knn}. 

The light quarks dynamics is more intricate because, as lattice simulations demonstrate, the quark sector (in the Landau gauge) becomes strongly coupled at infrared momenta \cite{Sternbeck:2017ntv,Kizilersu:2021jen} (no Landau pole is observed however). Remarkably, one-loop calculations in the CF model also exhibit the increase of the quark-gluon coupling in the infrared relative to the pure gauge coupling \cite{Pelaez:2015tba}. This suggests the self-consistent picture of a strongly interacting quark sector coupled to a perturbative gauge sector. In a recent article \cite{Pelaez:2017bhh}, we have proposed a systematic expansion scheme in the infrared regime based on a perturbative treatment of the pure gauge coupling together with an expansion in the inverse number of colors, $1/N_c$. At leading order, this results in the well-known rainbow-ladder resummation in the quark sector \cite{Johnson:1964da,Maskawa:1974vs,Maskawa:1975hx,Miransky:1984ef,Atkinson:1988mv,Atkinson:1988mw,Maris:1997hd,Maris:1997tm,Maris:2003vk,Bhagwat:2004hn,Roberts:2007jh,Eichmann:2008ae}, which correctly captures the essential aspects of dynamical chiral symmetry breaking. The advantages of the proposed expansion scheme is, first, that the rainbow-ladder resummation is obtained in a controlled manner and, second, that one can systematically implement standard QFT tools, such as renormalization and renormalization group (RG) improvement. The rainbow resummation of the quark propagator has been implemented in this context in Ref.~\cite{Pelaez:2017bhh} using a simple model for the running quark-gluon coupling---a simplification which has been removed in Ref.~\cite{Pelaez:2020ups}, where we have implemented a complete treatment of the RG running at leading order in the ``rainbow-improved'' loop expansion. Our results for the quark mass function are in very good agreement with lattice simulations for all values of the (degenerate) quark mass. 

Although the state of the art technology for handling the light quark sector of QCD with continuum approaches goes far beyond the rainbow-ladder resummation (see, for instance, \cite{Roberts:1994dr,Roberts:2000aa,Alkofer:2008tt,Cardona:2016bsq,Vujinovic:2018nko}), our work  provides an important new aspect in that it identifies relevant small parameters that one can use to obtain various levels of approximations in a systematic and controlled manner with, in principle, no need for ad-hoc parametrizations of the gluon propagator or the quark-gluon vertex. It is thus of interest to investigate to what extent our approach is able to describe other aspects of the light quark sector. One important application concerns the study of hadronic observables, which we undertake in the present work. In particular, we aim here at computing the prediction of the CF model for the pion decay constant $f_\pi$ at leading order in the rainbow-improved loop expansion, where the pion bound state corresponds to the resummation of ladder diagram with one-(massive)-gluon exchange. 

As a technical simplification, we shall compute $f_\pi$ in the chiral limit $m_\pi^2\to0$, whose value can be accurately deduced from the actual physical value with chiral perturbation theory at two-loop order \cite{Colangelo:2003hf}: $f_\pi(m_\pi^2=0)\approx86$~MeV. This presents important advantages. First, this allows us to use a small momentum expansion of Euclidean quantities (without the need for analytical continuation to Minkowski momenta) and, second, we can reduce the bound state problem to a set of coupled one-dimensional integral equations, allowing for a rather transparent and simple implementation of RG improvement---essential to correctly describe the UV tails---and for a simple numerical solution. We study the region of the (two-dimensional) parameter space for which $f_\pi$ equals its physical value, which allows us to fix in a physical manner the gluon mass parameter in terms of the coupling. The typical values we obtain are in agreement with previous results based on fitting lattice results for, say, the two-point functions in the Landau gauge. The present work is the first one where we constrain the parameter space using a physically measured quantity.

The article is organized as follows. Section \ref{sec_model} reviews the essentials of the rainbow improved expansion scheme at leading order. In Sec.~\ref{sec:fpi} we derive an exact expression for the decay constant $f_\pi(m_\pi^2=0)$ in terms of the Lorentz components of the quark propagator and of the components of the quark-antiquark-pion vertex in the limit of vanishing Euclidean pion momentum. The latter satisfy a set of coupled one-dimensional linear integral (Bethe-Salpether) equations derived in Sec.~\ref{sec:BSE}. At the present order of approximation, the kernel of these integral equations---corresponding to a one-gluon exchange---can be computed analytically. The renormalization and RG improvement of these integral equations is discussed in Sec.~\ref{sec:RG} and the ultraviolet behavior of the solutions is analyzed in Sec.~\ref{sec:UV}. The numerical solution of the BS equations and our results for  $f_\pi(m_\pi^2=0)$ in terms of the parameters of the model (the gluon mass and the quark-gluon coupling) is detailed in Sec.~\ref{sec:results}. In Sec.~\ref{sec:conclu} we discuss the conclusions and perspectives of the work. Finally, a series of technical details and results are gathered in Appendices \ref{Appsec:Ward}--\ref{Appsec:condensate}.

\section{The rainbow-improved loop expansion}
\label{sec_model}

We work with the Euclidean QCD action in the Landau gauge, supplemented with a gluon mass term
\begin{equation}
  \label{eq_action}
  \begin{split}
      S=\int d^4&x\Bigg[\frac 14 F_{\mu\nu}^aF_{\mu\nu}^a+ih^a\partial_\mu
    A_\mu^a +\partial_\mu\cb^a(D_\mu c)^a\\
&+\frac 12 m_\Lambda^2 (A_\mu^a)^2
   + \sum_{i=1}^{N_f}\bar\psi_i(\slashed D + {\cal M}_\Lambda)\psi_i  \Bigg].
  \end{split}
\end{equation}
Here, $F_{\mu\nu}^a=\partial_\mu A_\nu^a -\partial_\nu A_\mu^a+ g_\Lambda
f^{abc} A_\mu^b A_\nu ^c$ is the field-strength tensor and the covariant derivative is defined as $D_{\mu}X=\partial_{\mu}X -ig_\Lambda A_{\mu} X$, 
with $A_\mu$ the matrix gauge field in the appropriate representation. Also, $\slashed D=\gamma_\mu D_\mu$, where the
Euclidean Dirac matrices are chosen Hermitian and satisfy
$\{\gamma_\mu,\gamma_\nu\}=2\delta_{\mu\nu}$. Finally, the parameters $g_\Lambda$,
${\cal M}_\Lambda$ and $m_\Lambda$ are the bare coupling
constant, quark mass, and gluon mass, respectively, defined at some ultraviolet regulator
scale $\Lambda$. In the present paper, we are interested in the pion properties in the chiral limit and therefore, we only consider the case of $N_f=2$ degenerate quark
flavors. 

Thanks to the gluon mass term, the---otherwise standard---action \eqn{eq_action} possesses a well-defined perturbative expansion down to infrared scales. In perturbative calculations, this mass appears only in the bare gluon propagator,  $G_{\mu\nu}^{ab}(p)=\delta^{ab}G_{\mu\nu}(p)$
\begin{equation}
  \label{eq_gluon_propag}
  G_{\mu\nu}(p)=\frac{1}{p^2+m_\Lambda^2}\left(\delta_{\mu\nu}-\frac{p_\mu p_\nu}{p^2}\right).
\end{equation}
The latter is not modified at leading order in the RI loop expansion. Instead, the quark propagator gets dressed by rainbow diagrams and assumes the general form
\beq
\label{eq:quarkprop}
 S(p)=Z(p^2)\frac{i\slashed p+M(p^2)}{p^2+M^2(p^2)},
\eeq
where the quark field strength and mass functions assume the tree-level values $Z=1$ and $M={\cal M}_\Lambda$. Finally, the main quantity of interest in the present work is the quark-antiquark-pion vertex $\Gamma_\pi^i(q,q';p)$, where $i$ is the pion isospin index and where $q$ and $-q'$ denote the outgoing quark and antiquark momenta, whereas $p=q-q'$ is the incoming pion momentum. Here, the composite pion field is defined as $\pi^i(x)=\bar\psi(x)i\gamma_5\sigma^i\psi(x)$ (properly renormalized), with $\sigma^i$ the Pauli matrices in flavor space and $\gamma_5=\gamma_1\gamma_2\gamma_3\gamma_4$ such that $\gamma_5^\dagger=\gamma_5$ and $\{\gamma_5,\gamma_\mu\}=0$. Dirac, color, and flavor indices are left implicit.

The RI loop expansion relies on treating both the coupling in the pure gauge (ghost-gluon) sector $g_g$ and the inverse number of colors $1/N_c$ as small parameters, while keeping the quark gluon coupling $g_q$ arbitrary. \footnote{Strictly speaking, we need $g_q$ to remain, at most, of order one.} In practice, given a vertex function with $E$ external (quark, gluon or ghost) legs, the RI $\ell$-loop order is obtained as follows. Include first all standard diagrams up to $\ell$ loop. Each $\ell$-loop diagram involves given numbers of quark-gluon and pure gauge vertices and a given power of $N_c$ resulting from the color algebra. In terms of the rescaled (t'Hooft) couplings\footnote{The t'Hooft couplings are to be held fixed while taking the large $N_c$ limit.} $\hat g = g \sqrt{N_c}$, it scales as $\hat g_q^{k}\hat g_g^{k'}/N_c^p$, with $k+k'=2\ell+E-2$ and $0\le k\le2 \ell+E-2$, and where $p\ge E/2-1$. The rule is then to include as well and resum all higher loop diagrams with arbitrarily more quark-gluon vertices but with the same order in $\hat g_g$ and in $1/N_c$, that is, all $(\ell+n)$-loop diagrams of order $\hat g_q^{k+n}\hat g_g^{k'}/N_c^p$, with $n\ge0$. In particular, this systematically includes, at least, dressing the quark lines with the infinite series of rainbow diagrams, which are all $\sim \hat g_q^{k\ge0}(\hat g_g/N_c)^0$ and, hence, of the same order as the tree-level quark propagator. Moreover, it allows one to reproduce the correct perturbative behaviour in the in ultraviolet regime.
 
\begin{figure}[t]
 \includegraphics[width=.45\textwidth]{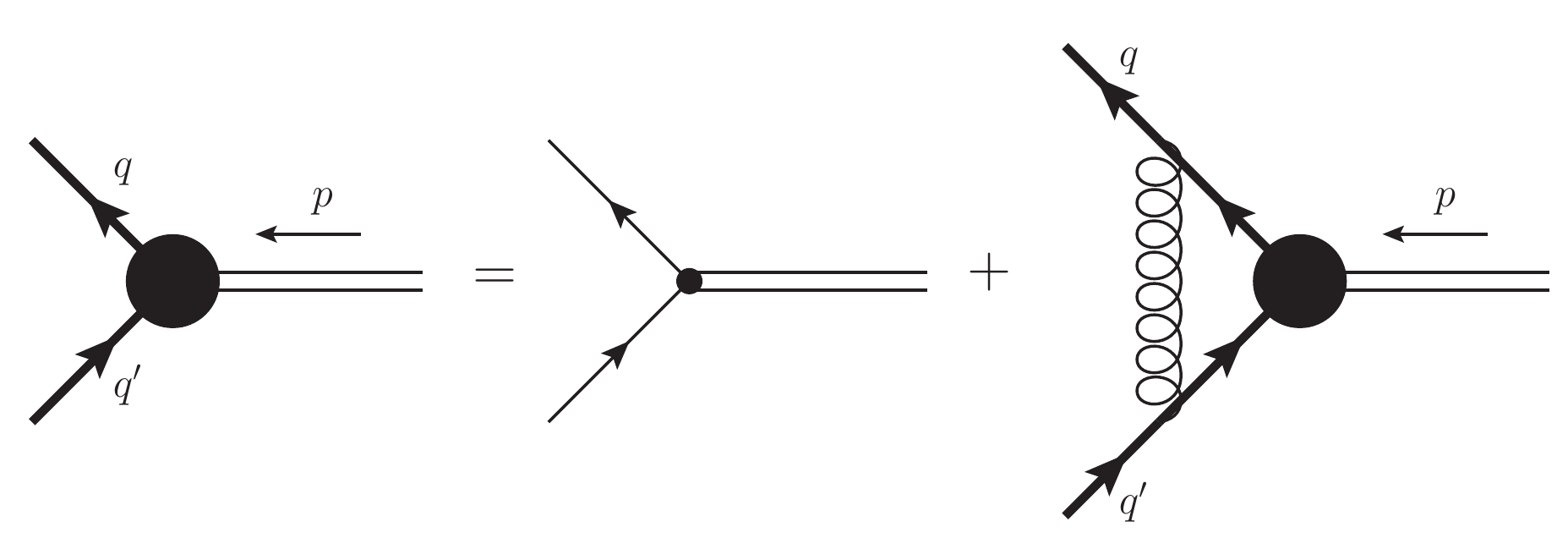}
 \caption{\label{Fig:BSE} The integral equation for the pion-quark-antiquark vertex function (black disk) at leading order in the RI-expansion: This generates the infinite series of ladder diagrams with rungs given by the tree level gluon propagator and quark-antiquark-gluon vertices and sides given by the leading-order (rainbow-resummed) quark propagator. The first term on the right-hand side is the tree-level vertex $i\gamma_5\sigma^i$. The present ladder resummation is directly relatedto the rainbow resummation for the quark propagator through the chiral Ward identities (see the Appendix \ref{Appsec:Ward}).}
\end{figure}

As detailed in Ref.~\cite{Pelaez:2017bhh}, at leading order, the tree-level gluon propagator \eqn{eq_gluon_propag} does not receive any correction. In contrast, as just explained, the whole series of rainbow diagrams contributes at the same order as the tree-level quark propagator and is thus to be resummed as the leading order in the RI loop expansion. Similarly, the infinite series of one-gluon exchange ladder diagrams with dressed quark lines contributes at the same order $\sim(\hat g_g/N_c)^0$ as the tree-level value to the quark-antiquark-pion vertex and thus constitute the RI leading order. The fact that both resummations come along is a manifestation of the axial Ward identities (see Appendix~\ref{Appsec:Ward}), which are thus consistently satisfied at this order of approximation. These resummations can be formulated in terms of the integral equations represented diagrammatically in Figs.~\ref{Fig:eq_rainbow} and \ref{Fig:BSE}. The quark propagator equation has been studied in detail in Refs.~\cite{Pelaez:2017bhh,Pelaez:2020ups}, to which we refer the reader for details. The equation for the pion vertex is the central focus of the present work. 
\begin{figure}[t!]
\includegraphics[width=.9\columnwidth]{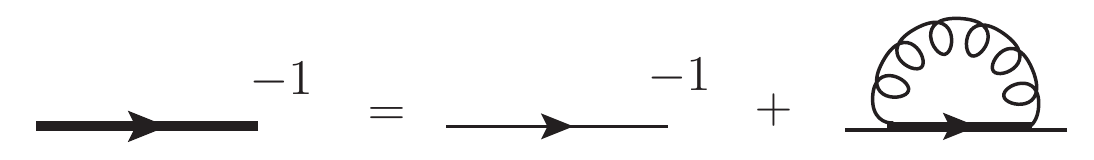},
 \caption{\label{Fig:eq_rainbow} The integral equation for the quark propagator (thick line) at leading order in the RI-expansion: This generates the infinite series of rainbow diagrams in terms of the tree-level propagators (thin lines) and vertices.} 
  \label{Fig:eq_rainbow}
\end{figure}

\section{Pion decay constant in the chiral limit}
\label{sec:fpi}

\begin{figure}[t!]
 \includegraphics[width=0.45\textwidth]{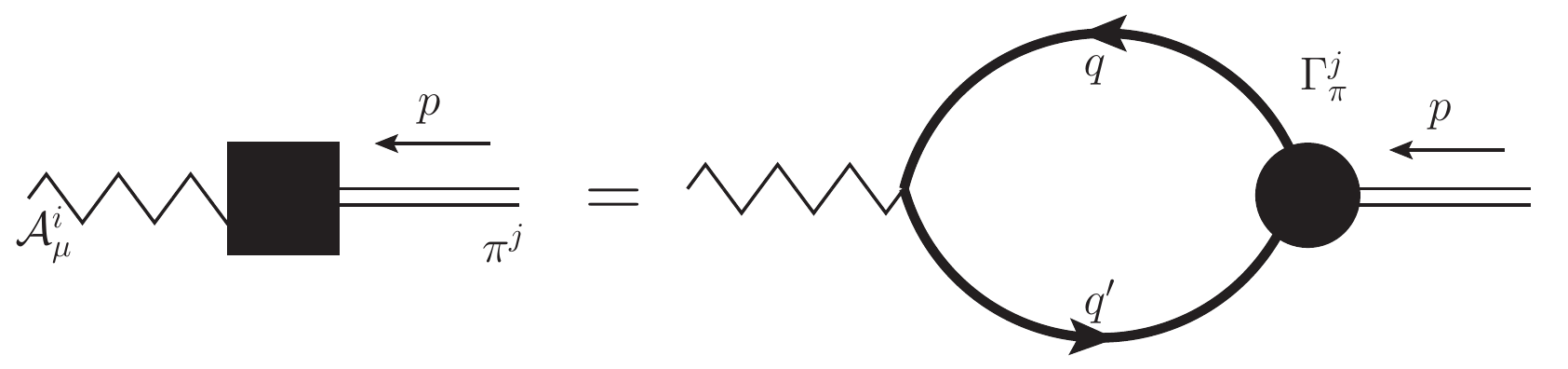}
 \caption{\label{Fig:propAPion} The ${\cal A}_\mu-\pi$ correlator (black square) in momentum space in terms of the quark propagator and of the pion-quark-antiquark vertex. The vertex on the left is the tree-level one: $i\gamma_\mu\gamma_5\sigma^i$. A similar expression with the dressed axial current-quark-antiquark vertex on the left and the bare pion--quark-antiquark one ($i\gamma_5\sigma^i$) on the right holds.}
\end{figure}

Let us first specify some conventions and normalizations. As recalled in Appendix \ref{Appsec:Ward}, the pion decay constant $f_\pi$ is related to the normalization of the axial current operator ${\cal A}_{\mu}^i(x)=\bar\psi(x)i\gamma_\mu\gamma_5\sigma^i\psi(x)$ and naturally appears in correlation functions involving the latter. For instance, the correlator $G_{{\cal A}_\mu\pi}^{ij}(x-y)=\langle {\cal A}_\mu^{i}(x)\pi^j(y)\rangle$ presents, in momentum space, a simple pole at the pion mass. With our choice of normalization (see Appendix \ref{Appsec:Ward}), we have, for $p^2\to-m_\pi^2$,
\beq
\label{eq:correlApi}
 G_{{\cal A}_\mu\pi}^{ij}(p)\sim-ip_\mu\delta^{ij}\frac{m_\pi^2}{{\cal M}_\Lambda}\frac{2f^2_\pi}{p^2+m_\pi^2}.
\eeq
The ratio $m_\pi^2/{\cal M}_\Lambda$, introduced here for convenience, remains finite and nonzero in the chiral limit, see \Eqn{appeq:GMOR}. The correlator \eqn{eq:correlApi} is related to the (bare) pion-quark-antiquark vertex $\Gamma_\pi^i(q,q')$ (see Fig.~\ref{Fig:propAPion} for conventions) as
\beq
 G^{ij}_{{\cal A}_\mu\pi}(p)=-\int_q \text{tr}\left[i\gamma_\mu\gamma_5\sigma^iS(q)\Gamma_\pi^j(q,q')S(q')\right]
\eeq
where the trace involves color, flavor, and Dirac indices and where $p=q-q'$.
Writing
\beq
\label{eq:vertexpole}
 \Gamma_\pi^i(q,q')=  i\gamma_5\sigma^i\frac{m_\pi^2}{{\cal M}_\Lambda}\frac{\gamma_\pi(q,q')}{p^2+m_\pi^2},
\eeq
where $\gamma_\pi(q,q')$ is regular when $p^2\to-m_\pi^2$, we deduce 
\beq
\label{eq:fpigen}
 -ip_\mu f_\pi^2=N_c\int_q \text{tr}\left[\gamma_\mu S(-q)\gamma_\pi(q,q')S(q')\right]_{p^2=-m_\pi^2},
\eeq
with a trace over Dirac indices. The symmetries of the problem (Lorentz, parity, charge conjugation invariance and  K-symmetry, see for instance \cite{Llewellyn-Smith:1969bcu}) constrain the Lorentz  structure of the pion vertex residues as
\begin{align}
\label{eq:lorentzdec}
\gamma_\pi(q,q')&=\gamma_P(q,q')+i\sigma_{\mu\nu}q_\mu q'_\nu\gamma_T(q,q')\nonumber\\
&+i\gamma_\mu\left[q_\mu\gamma_A(q,q')-q'_\mu\gamma_A(q',q)\right]  
\end{align}
with $\sigma_{\mu\nu}=\frac{i}{2}[\gamma_\mu,\gamma_\nu]$ and where $\gamma_{P,T,A}$ are real scalar functions. Furthermore, the functions $\gamma_P(q,q')$ and $\gamma_T(q,q')$ are symmetric under $q\leftrightarrow q'$. In general, one thus has to compute those scalar vertex functions, which depend on three scalar variables and which satisfy a set of coupled linear integral equations involving the resummed quark propagator (see Appendix~\ref{Appsec:BSE}).

In the chiral limit, where $m_\pi^2\to0$, the problem greatly simplifies and can be formulated in terms of three functions of one variable only. In particular, \Eqn{eq:fpigen} becomes
\beq
 -ip_\mu f_\pi^2=N_c\int_q \text{tr}\left[\gamma_\mu S(-q)\gamma_\pi(q,q')S(q')\right]_{p^2\to0} 
\eeq
and it is thus sufficient to expand  the right-hand side at linear order in $p$. Introducing the quark-antiquark relative momentum $r=(q+q')/2$, so that $q=r+p/2$ and $q'=r-p/2$, we have
\begin{align}
\label{eq:expPT}
 \gamma_{P,T}(q,q')&=\gamma_{P,T}(r^2)+{\cal O}(p^2)\\
\label{eq:expA}
 \gamma_A(q,q')&=\gamma_A(r^2)+\frac{p\cdot r}{r^2} \left[\gamma_A(r^2)-\gamma_B(r^2)\right]+{\cal O}(p^2),
\end{align}
where the RHSs define our notations. Expanding the quark propagator as well in \Eqn{eq:fpigen}, we have, in the chiral limit,
\begin{align}
\label{eq:fpibare}
f_\pi^2&= \frac{ N_c}{4\pi^2} \int_0^\infty \!\!dx \frac{xZ^2(x)}{\left[x+M^2(x)\right]^2}\Big\{\gamma_P(x)\!\left[M(x)-\frac{x}{2}M'(x)\right]\nonumber\\
&+\frac{3}{2} M(x)[x\gamma_T(x)-M(x)\gamma_A(x)]+ \frac{x+M^2(x)}{2}\gamma_B(x)\Big\}. 
\end{align}
As recalled in Appendix~\ref{Appsec:Ward}, the axial Ward identities imply that
\beq
\label{eq:Wardbare}
\gamma_P(x)=\frac{M(x)}{Z(x)}.
\eeq

Equation \eqn{eq:fpibare} is exact in the chiral limit. Retaining only the first line corresponds to the Pagel-Stokar approximation \cite{Pagels:1979hd,Roberts:1994dr}, which, thanks to the relation \eqref{eq:Wardbare}, provides an expression involving only the quark propagator. 
As we shall see below, the functions $\gamma_{T,A,B}(x)$ satisfy a set of coupled one-dimensional integral equations. These bare functions are to be renormalized and we shall discuss this issue together with the necessary RG improvement of the integral equations in the following. Note that \Eqn{eq:fpibare} implies that the integral on the RHS is finite and RG invariant.

\section{Bethe-Salpeter equation for the vertex}
\label{sec:BSE}

At leading order in the present expansion scheme, the quark-antiquark-pion vertex $\Gamma_\pi^i(q,q';p)$ resums the infinite series of ladder diagrams with rungs given by the tree-level one-gluon exchange \Eqn{eq_gluon_propag} and stiles given by rainbow-resummed quark propagators \Eqn{eq:quarkprop}; see \Fig{Fig:eq_rainbow}. This can be cast  into the linear integral equation depicted in \Fig{Fig:BSE}, which reads
\begin{align}
\label{eq:BSfullvertex}
\Gamma_\pi^i(q,q')=i\gamma_5\sigma^i-\lambda_\Lambda\!\int_k  G_{\mu\nu}(k)\gamma_\mu S(\ell)\Gamma_\pi^i(\ell,\ell')S(\ell')\gamma_\nu ,
\end{align}
where $\lambda_\Lambda=C_Fg_\Lambda^2$, with $C_F=(N_c^2-1)/(2N_c)\sim N_c/2$ for $N_c^2\gg 1$, $\ell=q-k$ and $\ell'=q'-k$.

Using the definition \eqn{eq:vertexpole} and the Lorentz decomposition \eqn{eq:lorentzdec}, one obtains a set of coupled integral equations for the scalar functions $ \gamma_{P,T,A,B}(q,q')$. Expanding the latter around $p^2=0$ up to linear order in $p^\mu$, this reduces to a set of one-dimensional integral equations for the functions $ \gamma_{P,T,A,B}(r^2)$ defined in Eqs.~\eqn{eq:expPT} and \eqn{eq:expA}. These equations are explicitly derived in the Appendix~\ref{Appsec:BSE}. 
The equation for $\gamma_P(r^2)$ actually decouples and reads, in the chiral limit
\beq
\label{Eq:smallPgammaP}
\gamma_P(r^2) =3\lambda_\Lambda\int_s \frac{Z^2(s^2)}{s^2 +M^2(s^2)}\frac{\gamma_P(s^2)}{(r-s)^2+m_\Lambda^2} .
\eeq
We recover the equation for the ratio $M(r^2)/Z(r^2)$ \cite{Maris:1997hd}, as expected from the Ward identity \eqn{eq:Wardbare}. 
The remaining equations read.
\begin{widetext}

\begin{align}
\label{Eq:smallPgammaT}
\gamma_T(x)&=\frac{\lambda_\Lambda}{16\pi^3}\int_0^\infty dy \left(\frac{x+y}{2x}f_{m_\Lambda^2}(x,y)
+\frac{(x-y)^2}{2 x}\Delta f_{m_\Lambda^2}(x,y)+\Delta I_{m_\Lambda^2}(x,y)\right)N(y)\\
\label{Eq:smallPgammaA}
\gamma_A(x)&=\frac{\lambda_\Lambda}{16 \pi^3}\int_0^\infty dy\Big\{\left[f_{m_\Lambda^2}(x,y)-\Delta I_{m_\Lambda^2}(x,y)\right] H(y)-\left[2 I_{m_\Lambda^2}(x,y)+(x-y)\Delta I_{m_\Lambda^2}(x,y)\right]L(y)\Big\}\\
\label{Eq:smallPgammaB}
\gamma_B(x)&=-\frac{3\lambda_\Lambda}{16 \pi^3}\int_0^\infty dy\Big\{\Delta I_{m_\Lambda^2}(x,y) H(y)
-\left[yf_{m_\Lambda^2}(x,y)-2 I_{m_\Lambda^2}(x,y)+ y \Delta I_{m_\Lambda^2}(x,y)\right]L(y)\Big\},
\end{align}
where 
\begin{align}
\label{eq:Nbare}
N(x)&=\left[\frac{Z(x)}{x+M^2(x)}\right]^2\left\{\gamma_P(x)+[x-M^2(x)] \gamma_T(x)-2M(x)\gamma_A(x)\right\}\\
\label{eq:Mbare}
H(x)&= \left[\frac{Z(x)}{x+M^2(x)}\right]^2\left\{M(x)\gamma_P(x)+2xM(x)\gamma_T(x)+\left[x-M^2(x)\right]\gamma_A(x)\right\}\\
\label{eq:Lbare}
L(x)&= \left[\frac{Z(x)}{x+M^2(x)}\right]^2\left\{ M'(x)\gamma_P(x)+M(x)\gamma_T(x)+\left[2+\frac{M^2(x)}{x}\right]\gamma_A(x)-\left[1+\frac{M^2(x)}{x}\right]\gamma_B(x)\right\},
\end{align}
\end{widetext}
and where we defined the functions
\begin{align}
\label{eq:fm}
f_{m^2}(x,y)&=\frac{\pi}{2x}  \left(b-\sqrt{b^2-xy}\right)\\
\label{eq:Im}
I_{m^2}(x,y)&=\frac{\pi }{2x}\left(by+\frac{2}{3x}\left[(b^2-xy)^{3/2}-b^3\right]\right),
\end{align}
with $b=(x+y+m^2)/2$, as well as 
\beq
 \Delta f_{m^2}(x,y)=\frac{f_{m^2}(x,y)-f_0(x,y)}{m^2}
\eeq
and similarly for $\Delta I_{m^2}(x,y)$.

\section{Renormalization and RG improvement }
 \label{sec:RG}
 
The above equations involve bare quantities and need to be properly renormalized. Also, a proper description of the ultraviolet regime requires to implement a RG improvement. As discussed in Ref.~\cite{Pelaez:2017bhh}, this also ensures to get a proper solution of the rainbow integral equation for the quark propagator. As we shall see below, this is also crucial so that the right-hand side of Eq.~\eqref{eq:fpibare} is finite. We refer the reader to Refs.~\cite{Pelaez:2017bhh,Pelaez:2020ups} for the treatment of the quark propagator equation and briefly recall the main necessary ingredients here.

We introduce the renormalized fields $A_{\mu}^a=\sqrt{Z_A}A_{R,\mu}^a$, $\psi=\sqrt{Z_\psi} \psi_R$, and $\bar\psi=\sqrt{Z_\psi} \bar\psi_R$, as well as the renormalized parameters\footnote{As mentioned before the quark-gluon and pure gauge couplings differ significantly in the infrared. It is, therefore, relevant to introduce different renormalization factors as $g_{\Lambda}=Z_{g_g} g_g=Z_{g_q} g_q$. This is discussed in detail in Ref.~\cite{Pelaez:2020ups} but will be of no direct relevance to the discussion below.} $m_\Lambda^2=Z_{m^2} m^2$, ${\cal M}_\Lambda=Z_{\cal M} {\cal M}$, and $g_{\Lambda}=Z_{g_q} g_q$. The renormalized quark propagator is obtained as 
\beq
 S(p)=Z_\psi(\mu_0^2)S_R(p;\mu_0^2),
 \eeq
where $\mu_0$ is an arbitrary renormalization scale. Clearly the mass function $M(p^2)$ is not renormalized, whereas we can define 
\beq
\label{eq:ZR}
  Z(p^2)=Z_\psi(\mu_0^2)Z_R(p^2;\mu_0^2).
 \eeq
 We choose the renormalization condition 
\beq
 Z_R(p^2=\mu_0^2;\mu_0^2)=1,
\eeq
 from which it follows, evaluating \Eqn{eq:ZR} at $\mu_0^2=p^2$, that
\beq
\label{eq:rencondZ}
 Z(p^2)=Z_\psi(p^2).
\eeq

We now come to the Bethe-Salpether equations \eqn{Eq:smallPgammaT}--\eqn{Eq:smallPgammaB}. These can be formally written as the following integral matrix equation
\beq
 \gamma(x)=g_\Lambda^2\int_0^\infty \!dy\, {\cal K}(x,y)\gamma(y),
\eeq
where $\gamma\equiv\left(\gamma_P,\gamma_T,\gamma_A,\gamma_B\right)$ and the matrix kernel ${\cal K}(x,y)$, which can be read off Eqs.~\eqn{Eq:smallPgammaT}--\eqn{eq:Lbare}, is proportional to one gluon propagator and two quark propagators, ${\cal K}\propto G S^2$; see \Eqn{eq:BSfullvertex}. Here, we made explicit the bare quark-gluon coupling constant but we leaved the gluon mass dependence of the kernel ${\cal K}$ implicit for the sake of the argument. We shall re-introduce it at the end. Upon introducting renormalized quantities as before as well as a renormalization factor $Z_\pi$ for the composite pion field, the renormalized kernel reads
\begin{align}
 \label{eq:renormkernel}
{\cal K}(x,y)&=Z_\psi^2(\mu_0^2)Z_A(\mu_0^2){\cal K}_R(x,y;\mu_0^2)
\end{align}
while the renormalized quark-antiquark-pion vertex is
\beq
\label{eq:Renormgamma}
\gamma(x)=Z_\psi^{-1}(\mu_0^2)Z_{\cal M}(\mu_0^2)Z_\pi(\mu_0^2)\gamma_R(x;\mu_0^2).
\eeq
The quark condensate operator $\sigma(x) = \bar\psi(x)\psi(x)$ is the chiral partner of the pion field and, thus receives the same renormalization factor: $Z_\sigma=Z_\pi$, Moreover, this operator is sourced by the tree-level quark mass ${\cal M}_\Lambda$, which implies that $Z_{\cal M}Z_\sigma=Z_{\cal M}Z_\pi$ is finite. Choosing a renormalization scheme with $Z_{\cal M}Z_\pi=1$, we have
\beq
\label{eq:renormvertex}
 \gamma(x)=Z_\psi^{-1}(\mu_0^2)\gamma_R(x;\mu_0^2).
\eeq

The equation for the renormalized vertex then reads
\begin{align}
 \gamma_R(x;\mu_0^2)&=Z_{g_q}^2(\mu_0^2)Z_\psi^2(\mu_0^2)Z_A(\mu_0^2)\times\nonumber\\
 & g_q^2(\mu_0^2)\int_0^\infty \!dy\, {\cal K}_R(x,y;\mu_0^2)\gamma_R(y;\mu_0^2).
\end{align}
As explained in Ref.~\cite{Pelaez:2020ups}, at the present order of approximation, we have $Z_{g_q}Z_\psi\sqrt{Z_A}= 1$, so that, finally,
\beq
\label{eq:renormBS}
 \gamma_R(x;\mu_0^2)=g_q^2(\mu_0^2)\int_0^\infty \!dy\, {\cal K}_R(x,y;\mu_0^2)\gamma_R(y;\mu_0^2)
\eeq
This equation is finite but involves potentially large logarithms, which can be resummed using renormalization group methods. First, we
can set $\mu_0^2=x$ in \eqn{eq:renormBS} to get
\beq
 \gamma_R(x;x)=g_q^2(x)\int_0^\infty \!dy\, {\cal K}_R(x,y;x)\gamma_R(y;x),
\eeq
with $g_q(x)$ the running quark-gluon coupling, to be determined from the appropriate beta function \cite{Pelaez:2020ups}. Then, we relate the functions ${\cal K}_R$ and $\gamma_R$ at different scales through Eqs. \eqn{eq:renormkernel} and \eqn{eq:renormvertex}:
\begin{align}
 \gamma_R(y;x)&={ \frac{Z_\psi^{-1}(y)}{Z_\psi^{-1}(x)}}\gamma_R(y;y)\\
 {\cal K}_R(x,y;x)&=\frac{Z_\psi^{2}(y)}{Z_\psi^{2}(x)}{\cal K}_R(x,y;y),
\end{align}
where we used the fact that, at the present order of approximation, the gluon propagator is at tree level so that $Z_A=1$. We also note that, at this order, $Z_\psi(x)$ is finite \cite{Pelaez:2017bhh}.
 
Defining $\hat\gamma(x)=Z_\psi(x)\gamma_R(x;x)$, we obtain the RG-improved equation
\beq
\label{RGimprovedBS}
 \hat\gamma(x)=g_q^2(x)\int_0^\infty \!dy\mathcal{K}_R(x,y;y)\hat\gamma(y),
\eeq
where the renormalized kernel ${\cal K}_R$ is computed as the bare one but with the bare quark and gluon propagators replaced by their renormalized counterpart at the scale $y$, that is with $Z(y)\to Z_R(y;y)=1$.
The previous argument is easily repeated to include the dependence of the kernel ${\cal K}$ on the gluon mass $m_\Lambda^2=Z_{m^2}(\mu_0^2)m^2(\mu_0^2)$, with $m^2(\mu_0^2)$ the renormalized square mass. At the present order of approximation, we have $Z_{m^2}=1$ and we conclude that, just as for the quark-gluon coupling, \Eqn{RGimprovedBS} involves the running gluon mass $m^2(x)$, obtained by solving the appropriate flow equation. The compete flow of the parameters $g_q(x)$ and $m^2(x)$ at leading order in the RI loop expansion has been discussed in Ref.~\cite{Pelaez:2020ups}, to which we refer the reader for details. Here we shall make direct use of the results presented there for these flows and for the RG-improved quark propagator.  We stress that, having a systematic set of expansion parameters allows us to  properly justify the RG improvement without any extra {\it ad hoc} hypothesis.

As clear from \Eqn{Eq:smallPgammaP}, the equation for $\hat\gamma_P(x)$ decouples from the others and we check that the resulting RG improved equation is consistent with the Ward identity for the renormalized vertex. In particular, the latter reads
\beq
\gamma_{R,P}(x;\mu_0^2)= \frac{M(x)}{Z_R(x;\mu_0^2)},
\eeq
from which it follows, using $Z_R(x;x)=1$, that
\beq
 \hat\gamma_P(x)=Z_\psi(x)M(x).
\eeq
One can check that the pseudoscalar component of the RG improved equation \eqn{RGimprovedBS} coincides with the RG improved equation for $M(x)$ obtained in Ref.~\cite{Pelaez:2020ups}. 

As a result, the equations for the remaining components $\hat\gamma_{T,A,B}(x)$ are linear integral equations with nonhomogeneous (source) terms given by the pseudoscalar contribution. Defining $\lambda(x)=C_Fg^2_q(x)$, these read, explicitly,
\begin{widetext}
\begin{align}
\label{eq:hatgammaT}
\hat \gamma_T(x)&=\frac{\lambda(x)}{16\pi^3}\int_0^\infty dy\left(\frac{x+y}{2x}f_{m^2}(x,y)
+\frac{(x-y)^2}{2 x}\Delta f_{m^2}(x,y)+\Delta I_{m^2}(x,y)\right)\hat N(y)\\
\label{eq:hatgammaA}
\hat\gamma_A(x)&=\frac{\lambda(x)}{16 \pi^3}\int_0^\infty dy\Big\{\left[f_{m^2}(x,y)- \Delta I_{m^2}(x,y)\right]\hat H(y)-\left[2 I_{m^2}(x,y)+(x-y)\Delta I_{m^2}(x,y)\right]\hat L(y)\Big\}\\
\label{eq:hatgammaB}
\hat\gamma_B(x)&=\frac{3\lambda(x)}{16 \pi^3}\int_0^\infty dy\Big\{-\Delta I_{m^2}(x,y)\hat H(y)
+\left[yf_{m^2}(x,y)-2 I_{m^2}(x,y)+ y \Delta I_{m^2}(x,y)\right]\hat L(y)\Big\},
\end{align}
where, as explained above, the gluon mass is the running one at the scale $x$, $m^2\equiv m^2(x)$ and where
\begin{align}
\label{eq:Nren}
\hat N(x)&=\hat N^{\rm source}(x)+\frac{[x-M^2(x)] \hat \gamma_T(x)-2M(x)\hat \gamma_A(x)}{\left[x+M^2(x)\right]^2}\\
\label{eq:Hren}
\hat H(x)&=\hat H^{\rm source}(x)+ \frac{2xM(x)\hat \gamma_T(x)+\left[x-M^2(x)\right]\hat \gamma_A(x)}{\left[x+M^2(x)\right]^2}\\
\label{eq:Lren}
\hat L(x)&=\hat L^{\rm source}(x)+ \frac{xM(x)\hat \gamma_T(x)+\left[2x+M^2(x)\right]\hat \gamma_A(x)-\left[x+M^2(x)\right]\hat \gamma_B(x)}{x\left[x+M^2(x)\right]^2},
\end{align}
\end{widetext}
with
\begin{align}
\label{eq:Nsource}
\hat N^{\rm source}(x)&=Z_\psi(x)\frac{M(x)}{\left[x+M^2(x)\right]^2}\\
\label{eq:Msource}
\hat H^{\rm source}(x)&= Z_\psi(x)\frac{M^2(x)}{\left[x+M^2(x)\right]^2}\\
\label{eq:Lsource}
\hat L^{\rm source}(x)&=Z_\psi(x)\frac{M(x)M'(x)}{\left[x+M^2(x)\right]^2}.
 \end{align}

Accordingly, we shall refer to the nonhomogeneous source terms $\hat\gamma_{T,A,B}^{\rm source}(x)$ as the right-hand-sides of Eqs.~\eqn{eq:hatgammaT}--\eqn{eq:hatgammaB}, with $\hat N(x)\to\hat N^{\rm source}(x)$, etc. These equations can be solved numerically, {\it e.g.}, by successive iterations of the source terms until convergence.

Finally, the pion decay constant in the chiral limit \Eqn{eq:fpibare} reads, in terms of renormalized quantities,
\begin{align}
\label{Eq:renFpi}
f_\pi^2&= \frac{ N_c}{4\pi^2}\! \int_0^\infty \frac{xdx}{{\left[x+M^2(x)\right]^2}}\Big\{ Z_\psi(x)M^2(x)\!\left[1-\frac{x}{2}\frac{M'(x)}{M(x)}\right]\nonumber\\
&+\frac{3}{2} M(x)[x\hat\gamma_T(x){-}M(x)\hat\gamma_A(x)]+ \frac{x+M^2(x)}{2}\hat\gamma_B(x)\!\Big\}.
\end{align}
We stress again that this equation is exact in the chiral limit. It reproduces Eq.(6.27) of Ref.~\cite{Roberts:1994dr} in the case were we only consider the pseudoscalar
tensor of the quark-pion vertex (first line), which is an extension of Pagels-Stokar formula \cite{Pagels:1979hd}.

For later use, we mention the following compact expression in terms of the functions \eqref{eq:Hren} and \eqref{eq:Lren}
\begin{align}
\label{Eq:renFpi2}
f_\pi^2&= \frac{ N_c}{4\pi^2} \int_0^\infty dxx\left[ \hat H(x)-\frac{x}{2} \hat L(x)\right].
\end{align}
The Pagel-Stokar formula then corresponds to keeping only the source terms \eqref{eq:Msource} and \eqref{eq:Lsource}.

\section{Ultraviolet behavior}
\label{sec:UV}

Before to present the numerical solution of the equations derived above, we analyze here the ultraviolet behavior of the solutions. We check explicitly that the integrals obtained by successive iterations of the source terms are ulraviolet convergent and we then solve for the leading large-momentum asymptotics of the vertex functions $\hat\gamma_{T,A,B}(x)$. The large momentum behaviors of the quark propagator in the chiral limit and of the quark-gluon coupling are\footnote{Of course, for dimensional reasons, the logarithmic terms must be understood as $\ln (x/x_0)$, with $x_0$ an arbitrary (though not too infrared) scale. We take $x_0=1$ for simplicity.} \cite{Roberts:1994dr,Pelaez:2020ups}
\beq
\label{eq:ZMUV}
Z_\psi(x)\sim1\,,\quad M(x)\sim\frac{A_M}{x}(\ln x)^{\gamma_M-1}\,, 
\eeq
and 
\beq
 \lambda(x)\sim\frac{C_F}{\beta_0\ln x},
\eeq 
with 
\beq
 \beta_0=\frac{11N_c-2N_f}{48\pi^2},
\eeq
and the quark mass anomalous dimension
\beq
\gamma_M=\frac{9C_F}{11N_c-2N_f}.
\eeq
The actual value of $\gamma_M$ is of importance in the following. In the large-$N_c$ limit used here, $\gamma_M=9/22\approx0.410$. For $N_c=3$ and $N_f=2$, $\gamma_M=12/29\approx0.414$. 

We then have the following leading ultraviolet behaviors for the functions \eqn{eq:Nsource}--\eqn{eq:Lsource}
\begin{align}
 \hat N^{\text{source}}(x)&\sim\frac{A_M}{x^3}(\ln x)^{\gamma_M-1}\\
 \hat H^{\text{source}}(x)&\sim \frac{A^2_M}{x^4}(\ln x)^{2\gamma_M-2}\\
\hat  L^{\text{source}}(x)&\sim -\frac{A^2_M}{x^5}(\ln x)^{2\gamma_M-2}.
\end{align}
Inserting these in Eqs.~\eqn{eq:hatgammaT}--\eqn{eq:hatgammaB} we obtain, for the source terms,
\begin{align}
\label{eq:TsourceUV}
 \hat \gamma_T^{\text{source}}(x)&\sim\frac{A_M}{12x^2}(\ln x)^{\gamma_M-1}\\
\label{eq:AsourceUV}
 \hat \gamma_A^{\text{source}}(x)&\sim \frac{c_A}{x\ln x}\\
\label{eq:BsourceUV}
  \hat \gamma_B^{\text{source}}(x)&\sim \frac{c_B}{x^2\ln x},
\end{align}
with
\begin{align}
\label{eq:cA}
c_A&=\frac{\gamma_M}{6}\int_0^\infty dxx\frac{Z_\psi(x)M^2(x)}{\left[x+M^2(x)\right]^2}\left[1-\frac{x}{2}\frac{M'(x)}{M(x)}\right]\\
\label{eq:cB}
c_B&=\frac{\gamma_M}{4}\int_0^\infty dxx^2\frac{Z_\psi(x)M^2(x)}{\left[x+M^2(x)\right]^2}\left[1-\frac{x}{3}\frac{M'(x)}{M(x)}\right].
\end{align}
 In deriving these asymptotic behaviors, we have used that, for $x\gg m^2$, the various functions in Eqs.~\eqn{eq:hatgammaT}--\eqn{eq:hatgammaB} read, for arbitrary $y$,
\begin{align}
\label{eq:f0}
 f_0(x,y)&=\frac{\pi}{2x}\left[y\theta(x-y)+(x\leftrightarrow y)\right]\\
  I_0(x,y)&=\frac{\pi y}{4x}\left[y\left(1-\frac{y}{3x}\right)\theta(x-y)+(x\leftrightarrow y)\right]\\
\Delta f_0(x,y)&=-\frac{\pi}{2x}\left[\frac{y}{x-y}\theta(x-y)+(x\leftrightarrow y)\right]\\
\label{eq:deltaI0}
 \Delta I_0(x,y)&=-\frac{\pi}{4x^2}\left[y^2\theta(x-y)+(x\leftrightarrow y)\right],
\end{align}
and, thus, in the same range of $x$,
 \begin{widetext}
\begin{align}
\label{eq:TUV}
\hat \gamma_T(x)&=\frac{\lambda(x)}{64\pi^2}\int_{0}^x dy\frac{y^2}{x^2}\hat N(y)+\frac{\lambda(x)}{64\pi^2}\int_x^\infty dy\hat N(y)\\
\label{eq:AUV}
\hat\gamma_A(x)&=\frac{\lambda(x)}{32 \pi^2}\int_{0}^x dy\left\{\frac{y}{x}\left[\hat H(y)-\frac{y}{2}\hat L(y)\right]+\frac{y^2}{2x^2}\left[\hat H(y)
-\frac{y}{3}\hat L(y)\right]\right\}+\frac{3\lambda(x)}{64 \pi^2}\int_x^\infty dy\left\{\hat H(y)+\left(\frac{5x}{9}-y\right)\hat L(y)\right\}\\
\label{eq:BUV}
\hat\gamma_B(x)&=\frac{3\lambda(x)}{64 \pi^2}\int_{0}^x dy\frac{y^2}{x^2}\left[\hat H(y)
-\frac{y}{3}\hat L(y)\right]+\frac{3\lambda(x)}{64 \pi^2}\int_x^\infty dy\left\{\hat H(y)
+\left(\frac{2x}{3}-y\right)\hat L(y)\right\}.
\end{align}
\end{widetext}

Writing $\hat\gamma=(\hat\gamma_T,\hat\gamma_A,\hat\gamma_B)$ and
\beq
\label{eq:iterativeeq}
 \hat\gamma=\hat\gamma^{\rm source}+\lambda\bar{\cal K}_R\cdot\hat\gamma,
\eeq
successive iterations of the source terms yield a formal expansion in powers of $\lambda(x)$
\beq
 \hat\gamma=\hat\gamma^{\rm source} + \sum_{n\ge1}\lambda^n\hat\gamma^{(n)}.
\eeq
One easily verifies that the first iteration of the source term yields, up to logarithms, $\hat\gamma^{(1)}_T\sim x^{-2}$ and $\hat\gamma^{(1)}_A\sim\hat\gamma^{(1)}_B\sim x^{-1}$ and that these power laws are stable against further iterations. Assuming that this is indeed the leading power-law behavior, we have
\begin{align}
\label{eq:NUV}
\hat N(x)&\sim\frac{M(x)}{x^2}+\frac{\hat \gamma_T(x)}{x}+{\cal O}\!\left(x^{-4}\right),\\
\label{eq:HUV}
\hat H(x)&\sim\frac{\hat \gamma_A(x)}{x}+{\cal O}\!\left(x^{-4}\right),\\
\label{eq:LUV}
\hat L(x)&\sim \frac{ 2\hat \gamma_A(x)-\hat \gamma_B(x)}{x^2}+{\cal O}\!\left(x^{-5}\right).
\end{align}

A detailed analysis of the leading ultraviolet behavior of the solutions is presented in Appendix~\ref{Appsec:UV}. We give here a brief summary. In all cases, the contributions $y\gg x$ to the integral equations are suppressed. The integrals in Eq.~\eqref{eq:TUV} are dominated by $y\sim x$, which yield contributions of the same order as the source term \eqref{eq:TsourceUV}. The equation for $\gamma_T$ decouples from those of $\gamma_A$ and $\gamma_B$ and is driven by the source term, that is, in turn, by the quark mass \eqref{eq:ZMUV}, with a modified coefficient $A_M/12\to A_M/11$ due from the integral contributions. Instead, the integrals in Eq.~\eqref{eq:AUV} are dominated by $y\ll x$, but, as before, these yield contributions of the same order as the source term \eqref{eq:AsourceUV}. It follows that $\gamma_A$ is also driven also driven by its source term with a modified coefficient $c_A\to \bar c_A$. Finally, the integrals in \eqref{eq:BUV} are dominated by $y\sim x$ and the source term \eqref{eq:BsourceUV} is subdominant. As a consequence $\gamma_B$ decouples (at leading order) and is driven by $\gamma_A\sim x^{-1}(\ln x)^{-1}$. We also find that the leading term $\sim x^{-1}(\ln x)^{-2}$ of each integral in Eq.~\eqref{eq:BUV} actually cancels out and that the resulting leading behavior of $\gamma_B$ is further suppressed by one inverse power of $\ln x$. The final result is 
 \begin{align}
 \label{eq:UVgammaT}
 \hat\gamma_T(x)&\sim\frac{A_M}{11x^2}(\ln x)^{\gamma_M-1},\\
 \label{eq:UVgammaA}
 \hat\gamma_A(x)&\sim\frac{\bar c_A}{x\ln x},\\ 
 \label{eq:UVgammaB}
 \hat\gamma_B(x)&\sim\frac{\gamma_M\bar c_A}{4x(\ln x)^3}.
 \end{align}
Using these behaviors, we show in Appendix~\ref{Appsec:UV} that the constant $\bar c_A$ verifies $\bar c_A=4\pi^2\gamma_Mf_\pi^2/(6N_c)$. Interestingly, we thus find that the UV asymptotics of both the pseudoscalar and the tensor components of the pion-quark-antiquark vertex is governed by the (renormalized) quark condensate $A_M\propto\langle\bar\Psi\Psi\rangle$ (see Appendix~\ref{Appsec:UVresults}) and the corresponding anomalous dimension $\gamma_M$, whereas that of the vector and pseudovector components is governed by $f_\pi^2$.
Finally, it is worth emphasizing that the enhanced logarithmic decay of $\hat\gamma_B$---with an exponent strictly larger than one---is crucial for the expression \eqn{Eq:renFpi} of $f_\pi$ to be finite. 

This last remark brings a question about how accurate the control of the UV tails must be to get a reliable determination of $f_\pi$. The UV contribution to Eq.~\eqref{Eq:renFpi} can be estimated as
\begin{align}
f_{\pi,{\rm UV}}^2&= \frac{ N_c}{4\pi^2} \int_{\Lambda^2}^\infty  dxx\left[H(x)-\frac{x}{2}L(x)\right]\\
&=\frac{N_c}{8\pi^2}\int_{\Lambda^2}^\infty dx \hat\gamma_B(x)+{\cal O}\left(A_M^2\Lambda^{-4}\right),
\end{align}
with $\Lambda$ a UV scale. Using the asymptotic behavior \eqref{eq:UVgammaB}, we deduce
\begin{equation}
\label{eq:fpiUV}
\frac{f_{\pi,{\rm UV}}^2}{f_\pi^2}=\frac{\gamma_M^2}{96(\ln \Lambda^2)^2}+{\cal O}\left(A_M^2\Lambda^{-4}\right).
\end{equation}
Despite the slow (logarithmic) convergence, the prefactor $\gamma_M^2/96\sim 10^{-3}$ ensures that this contribution is negligible.

\section{Results}
\label{sec:results}

In this section, we compute the pion decay constant in the chiral limit as a function of the parameters of the CF Lagrangian. We numerically solve Eqs.~(\ref{eq:hatgammaT}--\ref{eq:hatgammaB}) and we obtain $f_\pi$ from Eq.~(\ref{Eq:renFpi}). This requires prior knowledge of the running parameters $\lambda(x)$ and $m^2(x)$ and of the quark propagator functions $Z_\psi(x)$ and $M(x)$. We compute these quantities consistently within the present approximation scheme using the techniques put forward in Ref.~\cite{Pelaez:2020ups}. For completeness, we shall briefly recall the main aspects of the numerical procedure implemented there. 

At first, we need to set the scale of our calculation. We use the same procedure as in Ref.~\cite{Pelaez:2020ups}, which corresponds to fitting the quark propagator functions obtained by lattice simulations for physical values of the pion mass against the corresponding results in the present approach. In such a way our definition of the GeV corresponds to that of the lattice. 

As a first estimate, we can use the quark propagator functions obtained in this case---close to but not quite in the chiral limit---to compute the value of $f_\pi$ using the  expression derived above---valid in the chiral limit. This should provide a good estimate of the physical $f_\pi^{\rm phys}=92~{\rm MeV}$ as the chiral corrections are expected to be relatively small, roughly of the order of 5\%. We obtain\footnote{This corresponds to the parameters (see below): $g_0=1.94$, $m_0=0.15~{\rm GeV}$, and $M_0=3~{\rm MeV}$.} $f_\pi=87.9~{\rm MeV}$. For comparison the Pagel-Stockar approximation for this case gives $f_\pi^{\rm PS}=83.5~{\rm MeV}$.
 
As for our numerical procedure, we use a regular grid in the momentum $p=\sqrt x$ with a lattice spacing of $0.1\, {\rm GeV}$ divided in two regions. For momenta $p\le\Lambda_1=10\, {\rm GeV}$, we iterate the rainbow equations for the functions $Z_\psi(x)$ and $M(x)$ together with the corresponding RG equations for $\lambda(x)$ and $m^2(x)$ until convergence. As the integral rainbow equations involve integrating over large momenta, we use an extension of $Z_\psi(x)$ and $M(x)$ for $\Lambda_1\le p\le\Lambda_2=30\, {\rm GeV}$ determined by the UV expressions
\begin{align}
\label{UVexpression1}
Z_\psi^{\text{UV}}(x)&=1,\nonumber\\
M^{\text{UV}}(x)&=b_0\left(\ln \frac{x+m_0^2}{m_0^2}\right)^{\!\!-\gamma_M} +\frac{b_2}{x}\left(\ln \frac{x+m_0^2}{m_0^2}\right)^{ \gamma_M-1} .
\end{align}
\begin{figure}[t!]
 \includegraphics[width=0.45\textwidth]{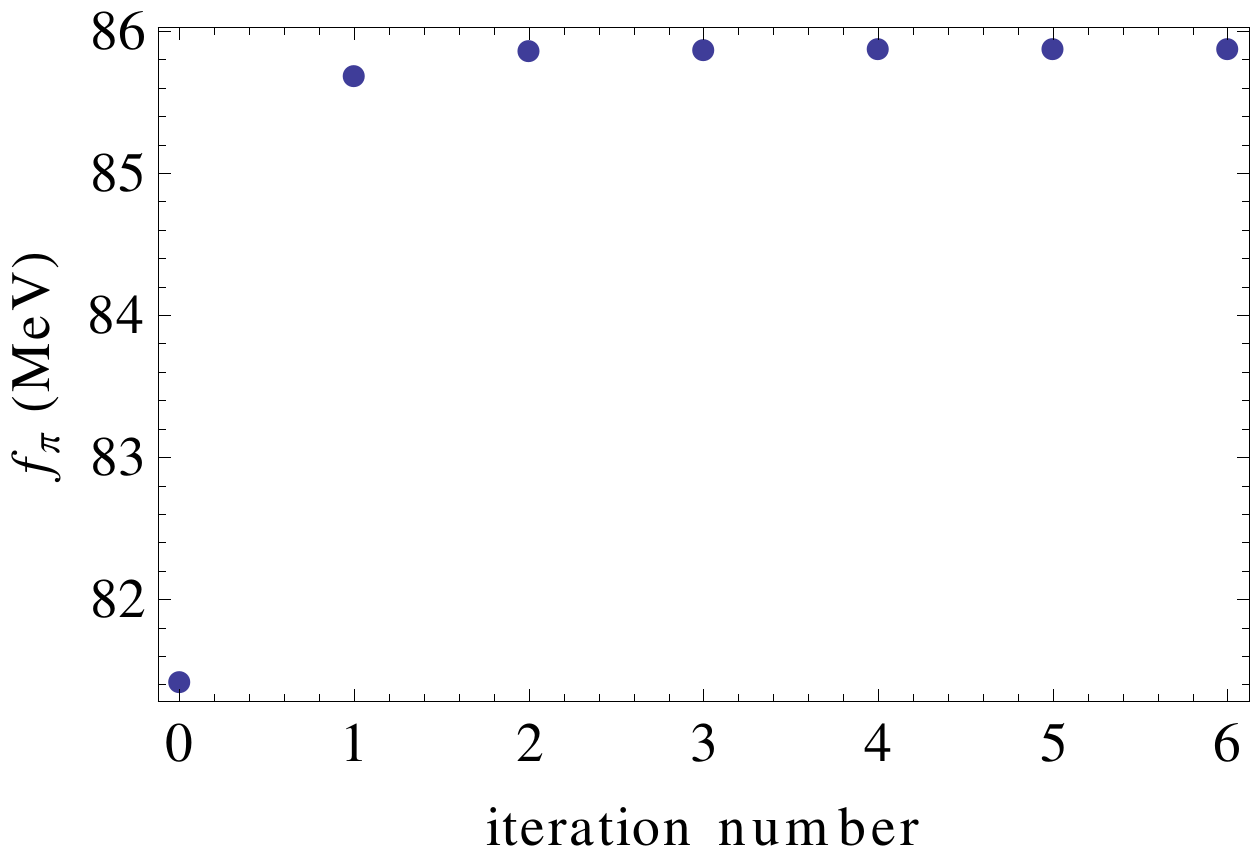}
 \caption{\label{Fig:fpi} Evolution of $f_\pi$ with the number of iterations for $g_0=1.93$ and $m_0=0.11$~GeV.}
\end{figure}
For the quark mass function we use a combination of the UV behaviors in either the chiral limit (term proportional to $b_2$) or the nonzero bare quark mass (term proportional to $b_0$). The coefficients, $b_0$ and $b_2$, are chosen in order to make $M(x)$ continuous and differentiable at $\Lambda_1$. 
The iteration starts with the functions \eqref{UVexpression1} extended to both regions and is done at fixed values of the input parameters $M_0=M(\Lambda_1^2)$, $m_0=m(\Lambda_1^2)$, and $\lambda_0=\lambda(\Lambda_1^2)$. The chiral limit is reached by lowering the value of $M_0$ until the contribution $\propto b_0$ in Eq.~\eqref{UVexpression1} becomes negligible over the whole range of momenta.\footnote{For instance, for $g_0=0.193$ and $m_0=0.11$~GeV, we have $b_0=4\times 10^{-4}~{\rm GeV}$ and $b_2= 0.227~{\rm GeV}^3$. We check that $b_0\ll b_2/\Lambda_1^2$.}
We use the lowest value for which our numerical algorithm is stable, that is, $M_0=0.5~{\rm MeV}$ and compute the functions $Z_\psi(x)$, $M(x)$, $\lambda(x)$, and $m(x)$ for various values of  $m_0$ and $\lambda_0$. In the following we quote the results in terms of the coupling $g_0=\sqrt{\lambda_0/C_F}$, with $C_F=4/3$.

\begin{figure}[t!]
\includegraphics[width=0.45\textwidth]{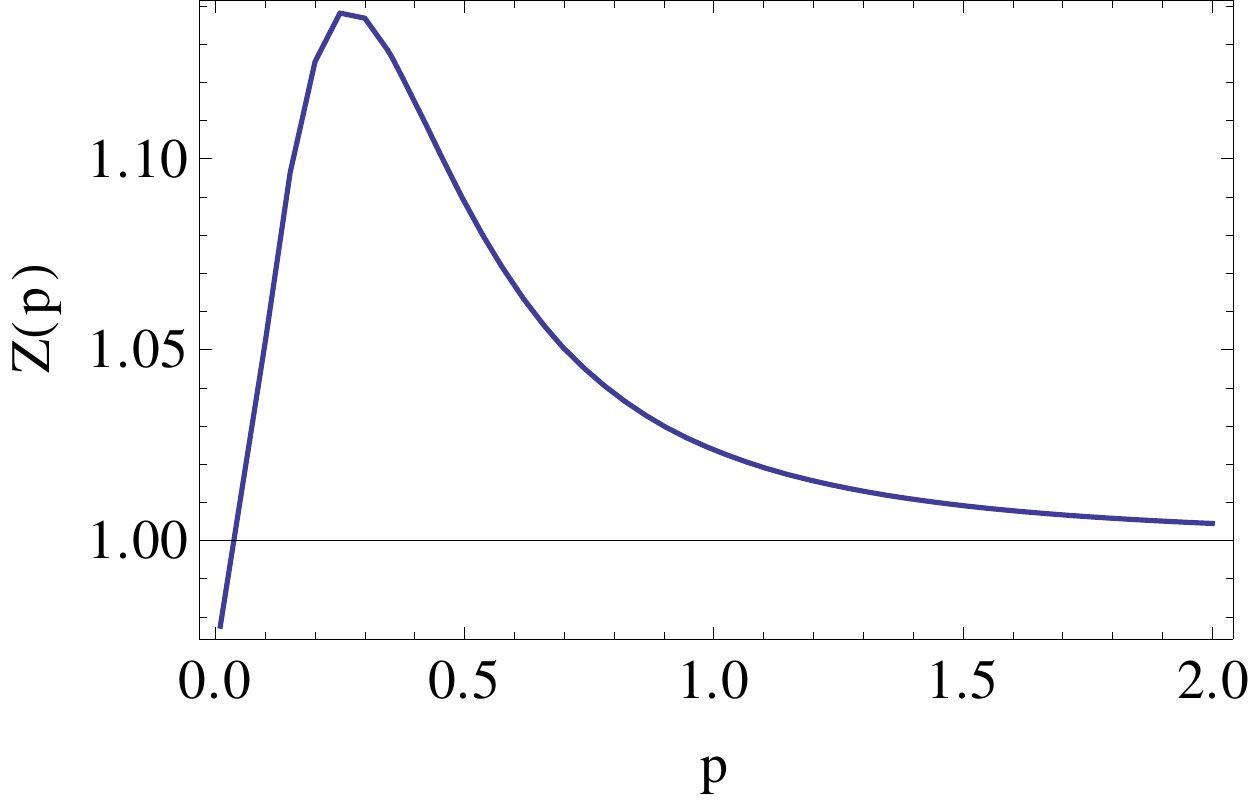}\\
\vspace{0.5cm}
\includegraphics[width=0.45\textwidth]{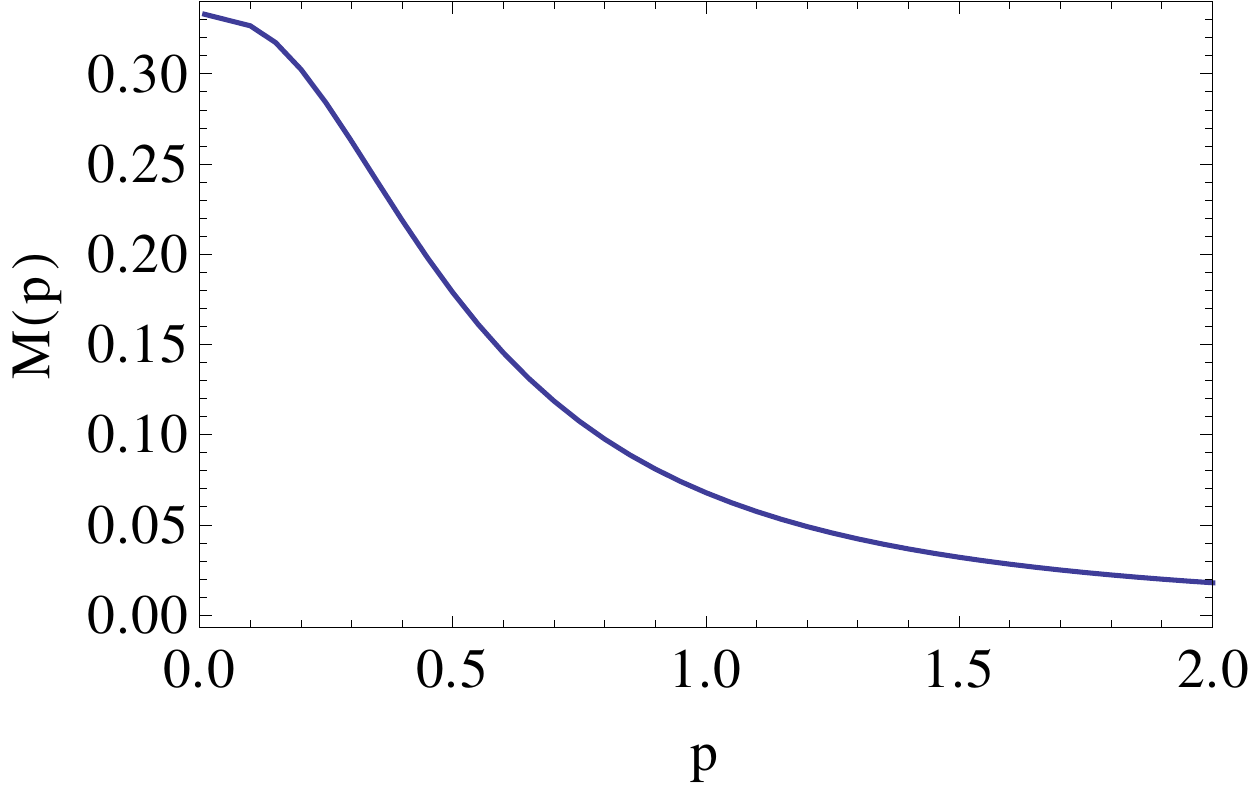}
  \caption{\label{Fig:propag} The quark propagator functions $Z_\psi$ and $M$ as functions of the momentum $p$ for $g_0=1.93$ and $m_0=0.11$ GeV. All units are in GeV.}
\end{figure}

\begin{figure}[t!]
\includegraphics[width=0.45\textwidth]{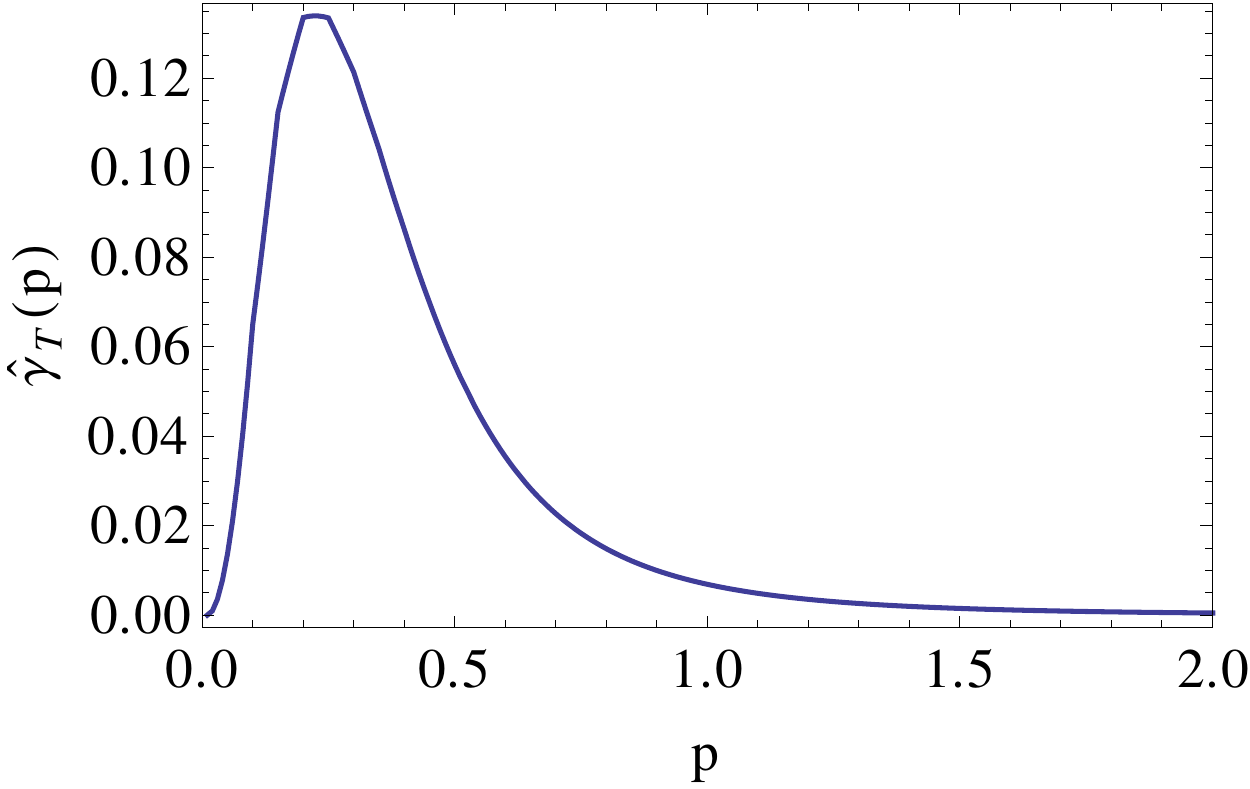}\\
\vspace{0.5cm}
\includegraphics[width=0.45\textwidth]{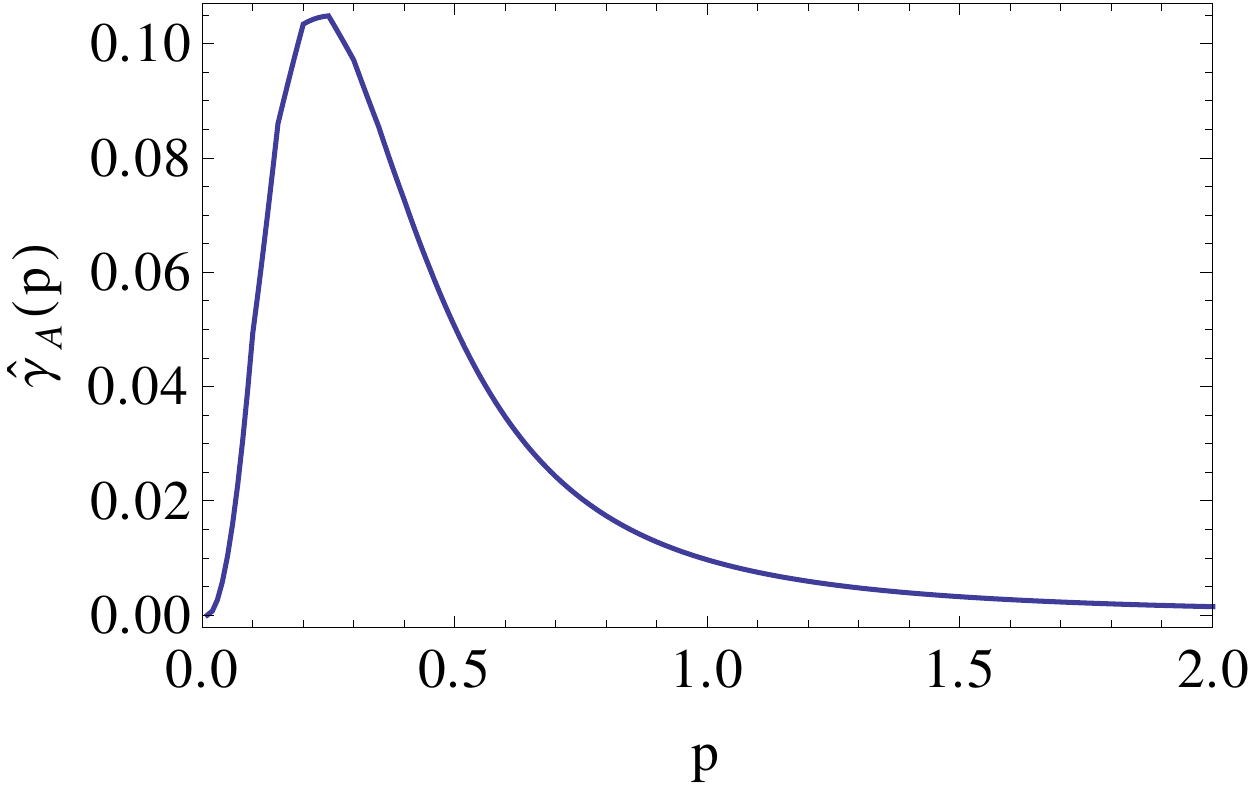}\\
\vspace{0.5cm}
\includegraphics[width=0.45\textwidth]{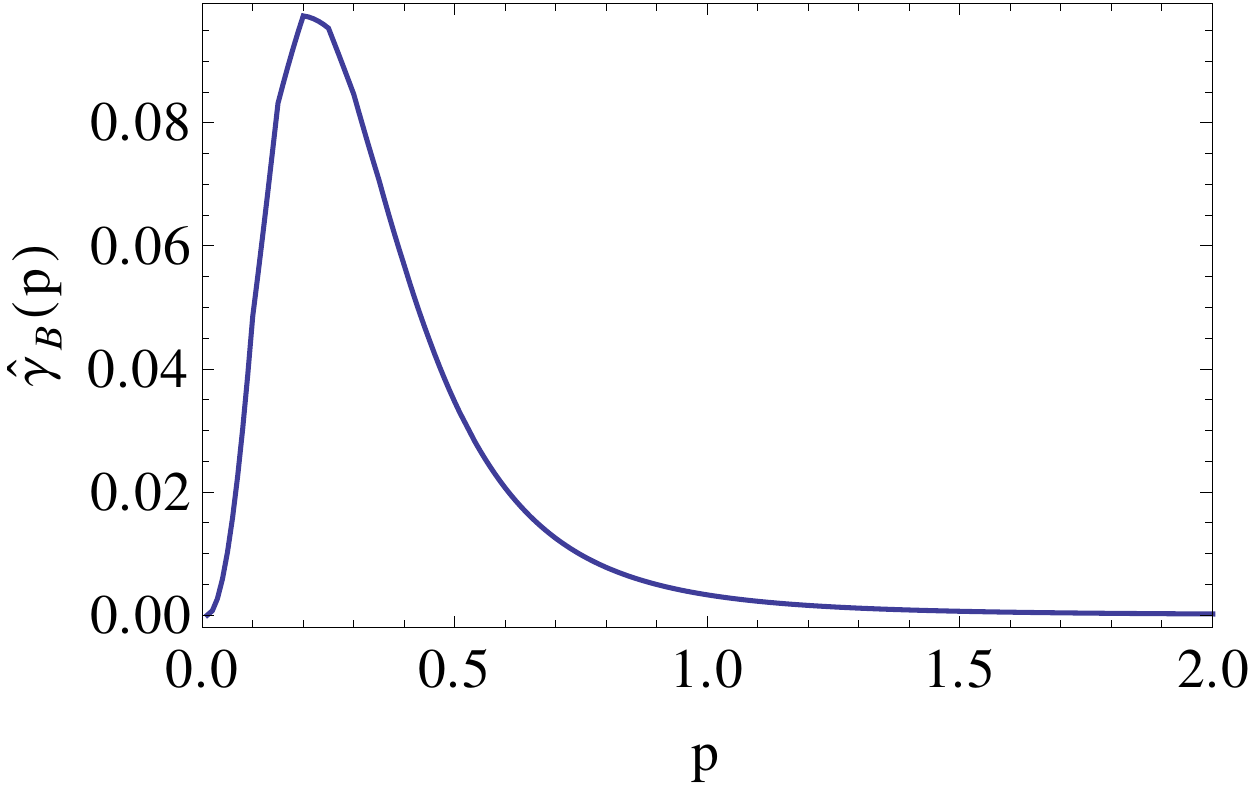}
  \caption{\label{Fig:gammas} Scalar functions $\hat\gamma_{T,A,B}$ as functions of the momentum $p$ for $g_0=1.93$ and $m_0=0.11$~GeV. All units are in GeV.}
\end{figure}

We then compute the pion vertex components $\hat\gamma_{T,A,B}(x)$, over the range $x\le\Lambda_1^2$ (in a grid in $p=\sqrt x$) by solving the system \eqref{eq:hatgammaT}--\eqref{eq:hatgammaB} recursively, with initial condition $\hat\gamma_T(x)=\hat\gamma_A(x)=\hat\gamma_B(x)=0$. The iterative process converges fast, typically after a few iterations only. This is illustrated in Fig.~\ref{Fig:fpi}, which shows the value of $f_\pi$ at each iteration for a typical choice of parameters. We see that the zeroth iteration, which corresponds to the Pagel-Stockar approximation, that is, which retains only the pseudoscalar component of the pion-quark-antiquark vertex, gives a relatively good approximation, $f_\pi^{(0)}=81.4~{\rm MeV}$, and that the tensor and vector components contribute about $5\%$ of the final $f_\pi=85.9~{\rm MeV}$ in that case. 
The (converged) functions $M$, $Z_\psi$, and $\hat\gamma_{T,A,B}$ for this set of parameters are shown in Figs.~\ref{Fig:propag} and \ref{Fig:gammas}. 

\begin{figure}[t!]
 \includegraphics[width=0.45\textwidth]{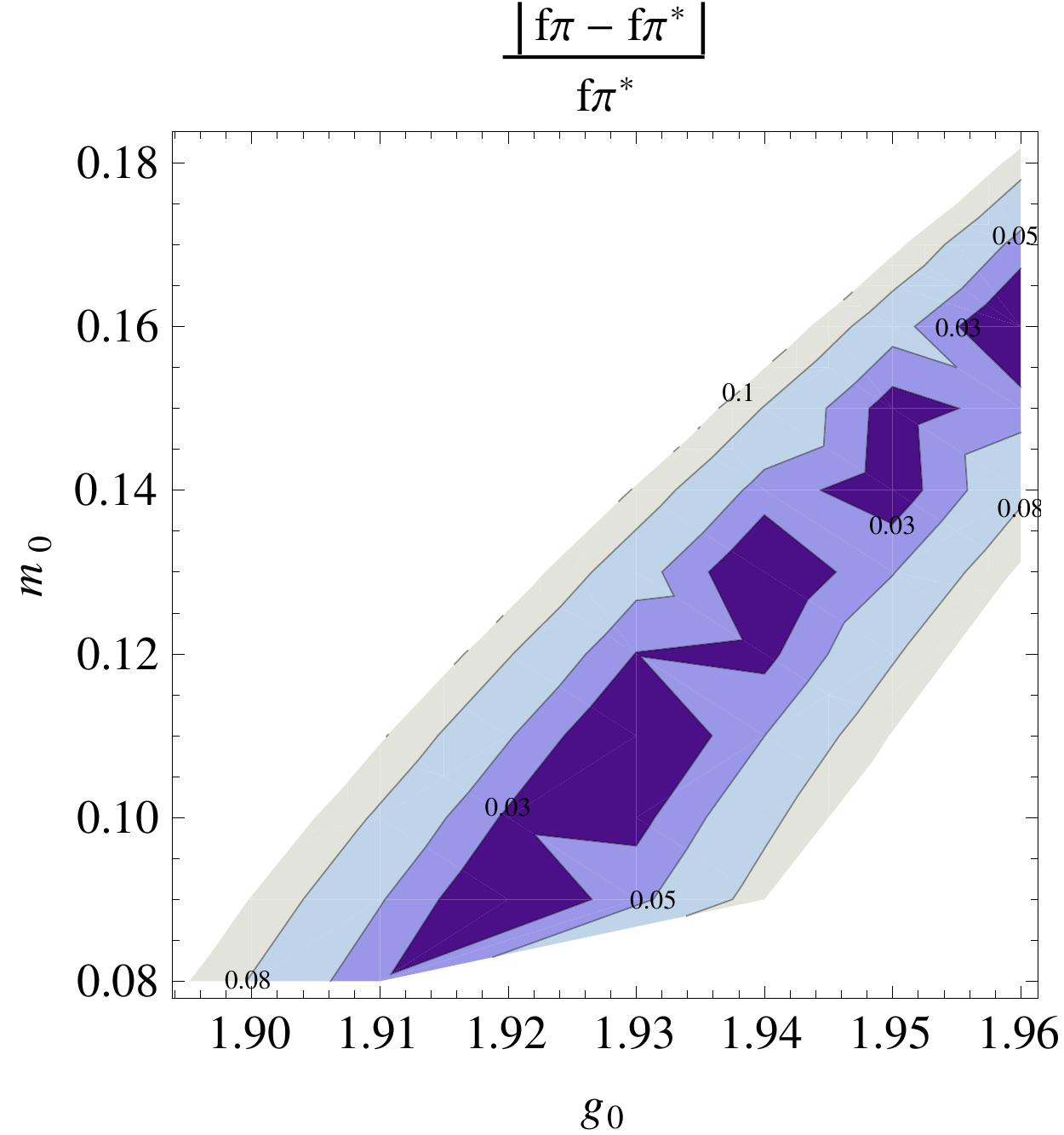}\\
\vspace{0.5cm}
 \includegraphics[width=0.45\textwidth]{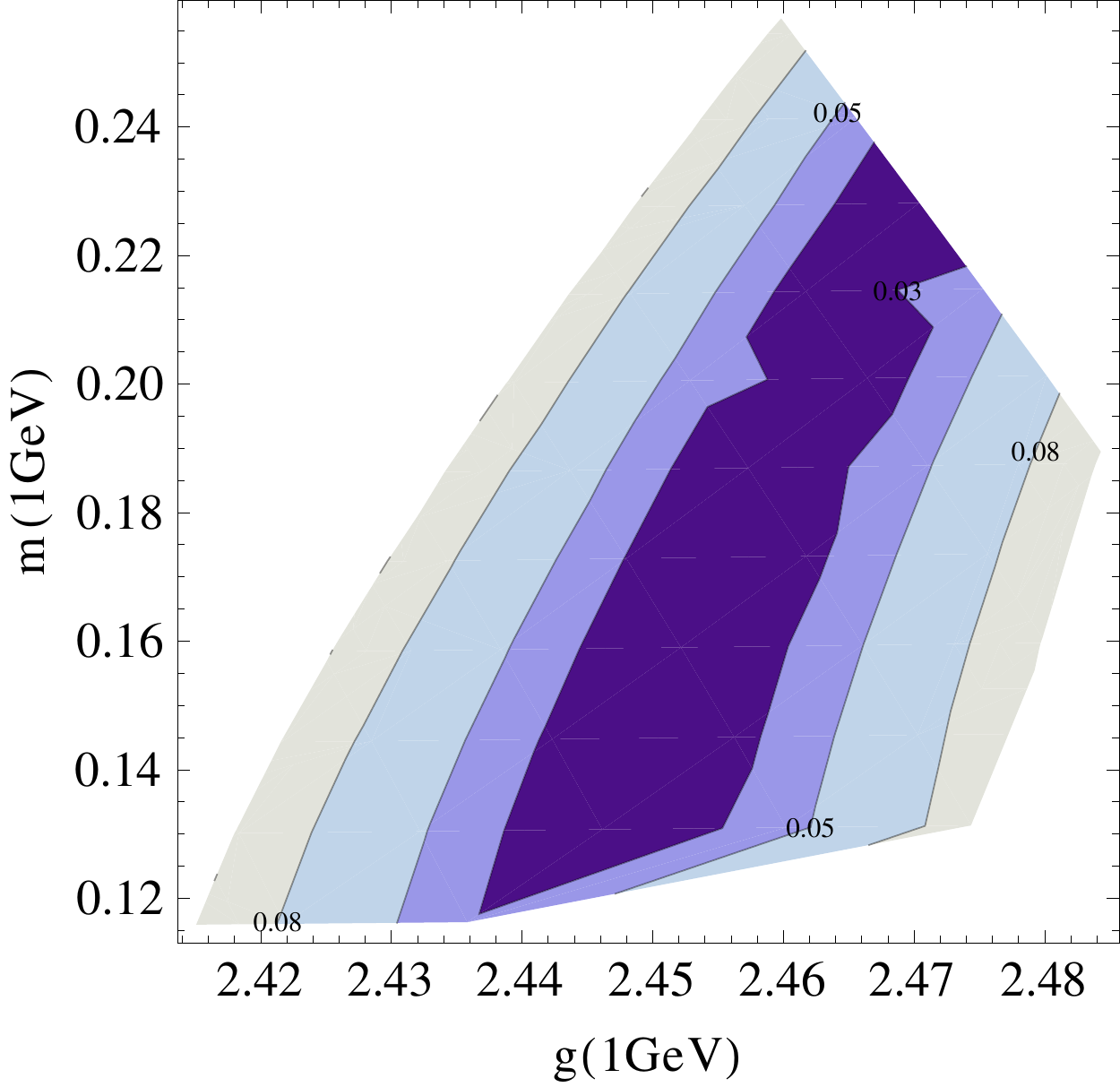}
 \caption{\label{Fig:fpimapeo} Regions in parameter space where $|f_\pi-f_\pi^*|/f_\pi^*$, with $f_\pi^*=86~{\rm MeV}$, is less than $3\%$, $5\%$, and $8\%$ (from darker to lighter) in terms of the parameter $m(\mu^2)$ and $g(\mu^2)$ at $\mu=10~{\rm GeV}$ (upper plot) and at $\mu=1~{\rm GeV}$ (lower plot).}
\end{figure}

\begin{figure}[t!]
 \includegraphics[width=0.45\textwidth]{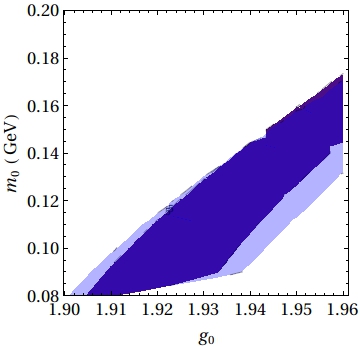}
 \caption{\label{Fig:MvsFpi} The region in parameter space where $|f_\pi-f_\pi^*|/f_\pi^*$ is less than $5\%$ (MeV (dark blue) compared to that where the overall error for the quark mass function compared with lattice data from \cite{Oliveira:2018lln} is less than $15\%$ (light blue).}
\end{figure}

Figure \ref{Fig:fpimapeo} shows the value of $f_\pi$ in the chiral limit as a function of the parameters $m_0$ and $g_0$. We also show the same plot in terms of the running parameters $m(\mu^2)$ and $g(\mu^2)$ evaluated at $\mu=1~{\rm GeV}$. The first main observation is that there exists values of these parameters for which $f_\pi$ is close to its physical value $f_\pi^*=86~{\rm MeV}$ in the chiral limit (deduced from the actual measured value by means of chiral perturbation theory \cite{Colangelo:2003hf}). The second important observation is that the parameters for which $f_\pi$ is close to its physical value are clearly correlated. Hence, fixing the value of $g_0$ essentially fixes the (physical) value of the mass parameter $m_0$. The physically acceptable values of the gluon mass parameter are then uniquely determined in terms of the coupling only. In particular, one can use these values to predict other quantities. As an immediate example here, we can compare the corresponding prediction for the quark mass function to the existing lattice results (in the chiral limit). We show in Fig.~\ref{Fig:MvsFpi} the values of the parameters for which the overall\footnote{The error functions are the ones defined in \cite{Pelaez:2020ups}} agreement between the predicted $M(x)$ and the lattice results of Ref.~\cite{Oliveira:2018lln} is less than $15\%$. This overlaps well with the region where $f_\pi$ is less than $5\%$ away from its expected value. 

\section{Summary and conclusions}
\label{sec:conclu}
We have computed the pion-quark-antiquark vertex function in the limit of vanishing pion momentum and the pion decay constant in the chiral limit in the context of the CF model approach to infrared QCD. The latter allows for a controlled expansion scheme in powers of both the coupling in the pure gauge sector and the inverse number of colors. At leading order, this leads to the resummation of rainbow-ladder diagrams in the quark sector with the tree-level (massive) gluon propagator and quark-gluon vertex. In the chiral limit, this results in a system of coupled one-dimensional integral equations for the various Lorentz components of the pion vertex. The RILO approximation allows us to implement the RG running of the parameters in a systematic and controlled manner, which is crucial in order to get consistent solutions and, in turn, a finite result for $f_\pi$. In particular this implies that there is no reliable solution in the limit $m_0\to0$ 
for which the RG running presents a Landau pole. 

We have obtained an exact expression for $f_\pi$ in the chiral limit that extends the known Pagel-Stockar approximation in terms of the vector and tensor components of the pion-quark-antiquark vertex. We have performed a detailed analysis of the UV behavior of the relevant functions with the interesting results that the power-law decays in the chiral limit are controlled by either the quark condensate or the pion decay constant. Finally we have obtained a numerical solution of the RG-improved coupled integral equations in terms of the parameters of the model, namely the gluon mass parameter $m_0$ and the coupling $g_0$. 

Our main result is that there exist correlated values of the parameters $m_0$ and $g_0$ for which the pion decay constant $f_\pi$ takes its physical value in the chiral limit. This thus defines a physical constraint $m_0^{\rm phys}(g_0)$ which allows one to predict other quantities in terms of the coupling only. Of course, it would be be extremely interesting to fit a second experimental observable to fully determine the two parameters directly from experimental data. 

One possibility would be to use the transition temperature associated to the QCD phase transition. Studies of the deconfinement transition exist within the CF model \cite{Reinosa:2014zta,Reinosa:2014ooa,Reinosa:2015gxn} but they have been so far restricted to the case of pure Yang-Mills theory or QCD in the limit where all quarks are considered heavy. Those situations are very far from the chiral limit addressed which prevents us from combining the results. For this reason, it becomes pressing to extend the study of the QCD phase structure within the CF model to the light quark region. Part of this analysis in under way.

A second quantity that could be used to fully determine the parameters of the CF model is the strong coupling constant $\alpha_S$. However, to make the comparison reliable it would be necessary to include two elements that are beyond the scope of the present study. First, one would need to establish the evolution of the coupling constant in a realistic way (including the various heavier quarks) up to the scales where the coupling $\alpha_S$ is small and well measured. In particular, this would require including two loop corrections to the running, which has already been done in the $N_f=2$ case, see \cite{Barrios:2021cks}, and could easily be extended above the heavier quark thresholds. Second, it would be necessary to establish, in the weak coupling regime, the correspondence between the running calculated here in the Taylor scheme with the $\bar{MS}$ which is the scheme usually reported in the literature.

Once the parameters are fully determined in that way, one could envisage studying the predictions of our approach for the pion bound state at nonzero pion mass or other light mesonic bound states. There exist well-developed techniques to study the relevant integral equations (see, for instance,\cite{Carbonell:2010zw,Fischer:2014xha,Eichmann:2016yit,Vujinovic:2018nko}) which could be easily implemented in the CF model. 

Beyond these considerations, we stress that another interesting take on the present work is that the CF model in fact provides a well-defined notion of a gluon mass parameter that could serve as a benchmark for testing the masslessness of the gluon. Giving reliable experimental constraints on the gluon mass requires a proper definition of the latter. The situation is similar to the case of the quark mass or of the gauge coupling, which being unphysical, require a proper definition ({\it e.g.} defined at a given scale in a given scheme) in order to be given experimental constraints/values. Although the latter is well understood and has been studied in great detail \cite{ParticleDataGroup:2022pth}, the theoretical status of the gluon mass is much less clear. For instance, the particle data book \cite{ParticleDataGroup:2022pth} mentions limits on a possible gluon mass that are based on ideas from the early days of QCD, which are now completely obsolete, in particular, because the notion of gluon mass used there is ill-defined. The CF model offers a proper theoretical definition of a gluon mass parameter that can be constrained by experimental data. The present work makes a step in that direction.

\begin{acknowledgments}

The authors would like to acknowledge the financial support from PEDECIBA program and from the  ANII-FCE-126412 and ANII-FCE-166479 project and from the CNRS-PICS project {\it irQCD}.
N.W. thanks the Universit\'e Paris Sorbonne, where part of this work has been realized, for hospitality. U.R. and J.S. acknowledge the support and hospitality of the Universidad de la Rep\'ublica de Montevideo during various stages of this work.
\end{acknowledgments}

\pagebreak

\appendix

\section{Axial Ward Identities and their consequences}
\label{Appsec:Ward}

Introducing source terms for the chiral multiplets $(\sigma,\pi^i)$ and $({\cal V}_\mu^i,{\cal A}_\mu^i)$, with the composite fields $\sigma(x)=\bar\psi(x)\psi(x)$, $\pi^i(x)=\bar\psi(x)i\gamma_5\sigma^i\psi(x)$, ${\cal V}_\mu^i(x)=\bar\psi(x)i\gamma_\mu\sigma^i\psi(x)$, and ${\cal A}_\mu^i(x)=\bar\psi(x)i\gamma_\mu\gamma_5\sigma^i\psi(x)$, the QCD action is modified as\footnote{Both the FP gauge-gixing terms and the CF gluon mass term in the action are insensitive to chiral transformation and do not alter the present discussion.}
\beq
\label{appeq:chiralsource}
 S_{\rm QCD}\to S_{\rm QCD}-S_{\rm s},
\eeq
with
\begin{align}
\label{appeq:chiralsourceterm}
 S_{\rm s}=\int d^4x \left\{\bar\eta\psi+\bar\psi\eta+J\sigma+J^i\pi^i+J_\mu^i{\cal V}_\mu^i+L_\mu^i{\cal A}_\mu^i\right\}.
\end{align}
Using the invariance of the functional integration measure under infinitesimal axial $SU_A(N_f)$ transformations of the quark fields, $\delta^i_\chi\psi=i\sigma^i\gamma_5\psi$ and $\delta^i_\chi\bar\psi=i\bar\psi\sigma^i\gamma_5$, one derives the following (Ward) identity in terms of the effective action $\Gamma[\psi,\bar\psi,{\cal J}]$ at nonzero  sources ${\cal J}=(J, J^i, J_\mu^i, L_\mu^i)$:
\begin{align}
\label{appeq:wardvertex}
&({\cal M}_\Lambda-J)\frac{\delta\Gamma}{\delta J^i}+J^i\frac{\delta\Gamma}{\delta J}-\epsilon^{ijk}\!\left[J_\mu^j\frac{\delta\Gamma}{\delta L_\mu^k}+L_\mu^j\frac{\delta\Gamma}{\delta J_\mu^k}\right]\nonumber\\
&-\frac{1}{2}\partial_\mu\frac{\delta\Gamma}{\delta L_\mu^i}+\bar\psi\frac{i\gamma_5\sigma^i}{2}\frac{\delta\Gamma}{\delta \bar\psi}-\frac{\delta\Gamma}{\delta \psi}\frac{i\gamma_5\sigma^i}{2}\psi=0,
\end{align}
where the first term in the last line stems from the fact that we considered gauged axial transformations. Taking functional derivatives and evaluating at vanishing sources yields the set of axial Ward identities relating various vertex and correlation functions. 

We first discuss the correlators
\begin{align}
 G^{ij}_{\pi\pi}(x-y)&=\langle\pi^i(x)\pi^j(y)\rangle=-\left.\frac{\delta^2\Gamma}{\delta J^i(x)\delta J^j(y)}\right|_{{\cal J}=0}\\
 G^{ij}_{{\cal A}_\mu\pi}(x-y)&=\langle{\cal A}_\mu^i(x)\pi^j(y)\rangle=-\left.\frac{\delta^2\Gamma}{\delta L_\mu^i(x) \delta J^j(y)}\right|_{{\cal J}=0}.
\end{align}
Eq.~\eqref{appeq:wardvertex} implies the following identity, in momentum space,
\beq
\label{appeq:Wardpropag}
 {\cal M}_\Lambda G^{ij}_{\pi\pi}(p)+i\frac{p_\mu}{2} G^{ij}_{{\cal A}_\mu\pi}(p)=-\delta^{ij}\sigma,
\eeq
where $\sigma=\langle \bar\psi(x)\psi(x)\rangle=-\int_q{\rm tr}S(q)$ is the quark condensate. With our convention, $p_\mu$ is the outgoing (incoming) axial vector (pion) momentum for the correlator $G^{ij}_{{\cal A}_\mu\pi}(p)$; see Fig.~\ref{Fig:propAPion}. 

The pion decay constant $f_\pi$ characterizes the amplitude of the pion-to-lepton disintegration and is related to the normalization of the axial vector operator\footnote{The amplitude of the matrix element of the axial vector operator between the hadronic vacuum $\ket{0}$ and on-shell one-pion states $\ket{\pi^i(\tilde p)}$, with the Minkowskian $4$-momentum $\tilde p^\mu=(\varepsilon_p,\vec p)$, where $\varepsilon_p=\sqrt{\vec p^2+m_\pi^2}$ is fixed by using Lorentz invariance and isospin symmetry. We write, with Lorentz-invariant normalizations of the one-particle states,
\begin{align}
\bra{0}\tilde \pi^i(\tilde x)\ket{\pi^j(\tilde p)}&=e^{-i\tilde p\cdot \tilde x}\delta^{ij}\sqrt{N_\pi}\nonumber\\
\bra{0}\tilde {\cal A}_{\mu}^i(\tilde x)\ket{\pi^j(\tilde p)}&=-i \tilde p_\mu e^{-i\tilde p\cdot \tilde x}\delta^{ij}2f_\pi\sqrt{N_{\cal A}},\nonumber
\end{align}
where the tildes refer to Minkowskian quantities.} ${\cal A}_\mu$. In the chiral limit, one has an isolated one-particle (pion) pole in the vicinity of $p^2=0$ and the propagators in \Eqn{appeq:Wardpropag} have the analytic structures
\begin{align}
\label{appeq:polepipi}
 G^{ij}_{\pi\pi}(p)&\sim\delta^{ij} \frac{N_\pi}{p^2+m_\pi^2}\\
\label{appeq:poleApi}
 G^{ij}_{{\cal A}_\mu\pi}(p)&\sim-ip_\mu\delta^{ij}\frac{2f_\pi\sqrt{N_{\cal A}N_\pi }}{p^2+m_\pi^2}
\end{align}
in a finite interval of $p^2$, where $N_\pi$ and $N_A$ are some normalization factors.\footnote{Note that we are dealing with bare fields and, in particular, $N_\pi$ is not to be confused with the renormalization factor $Z_\pi$ which defines the renormalized pion field in \Eqn{eq:Renormgamma}.} Writing the identity \eqn{appeq:Wardpropag} for $p^2\to-m_\pi^2$, we get the relation 
\beq
\label{appeq:Npi}
 {\cal M}_\Lambda \sqrt{N_\pi}=m_\pi^2f_\pi\sqrt{N_{\cal A}}.
\eeq
The fact that $Z_{\cal M}Z_\pi$ is finite implies that the product ${\cal M}_\Lambda \sqrt{N_\pi}$ is finite and, hence, $N_{\cal A}$ as well. The standard definition of $f_\pi$ \cite{Weinberg:1996kr} corresponds to choosing $N_{\cal A}=1$, from which we arrive at \Eqn{eq:correlApi}. Also, in the chiral limit, where $m_\pi^2\to0$, the expressions \eqn{appeq:polepipi} and \eqn{appeq:poleApi} are valid near $p=0$. Writing the identity \eqn{appeq:Wardpropag} at $p=0$ yields
\beq
 \frac{{\cal M}_\Lambda N_\pi}{m_\pi^2}=-\sigma,
\eeq
where $\sigma$ is the quark condensate in the chiral limit. Together with \Eqn{appeq:Npi}, this yields the famous Gell-Mann-Oaked-Renner relation \cite{Gell-Mann:1968hlm}
\beq
\label{appeq:GMOR}
-\sigma {\cal M}_\Lambda=f_\pi^2m_\pi^2.
\eeq 

\begin{figure}[t!]
 \includegraphics[width=0.45\textwidth]{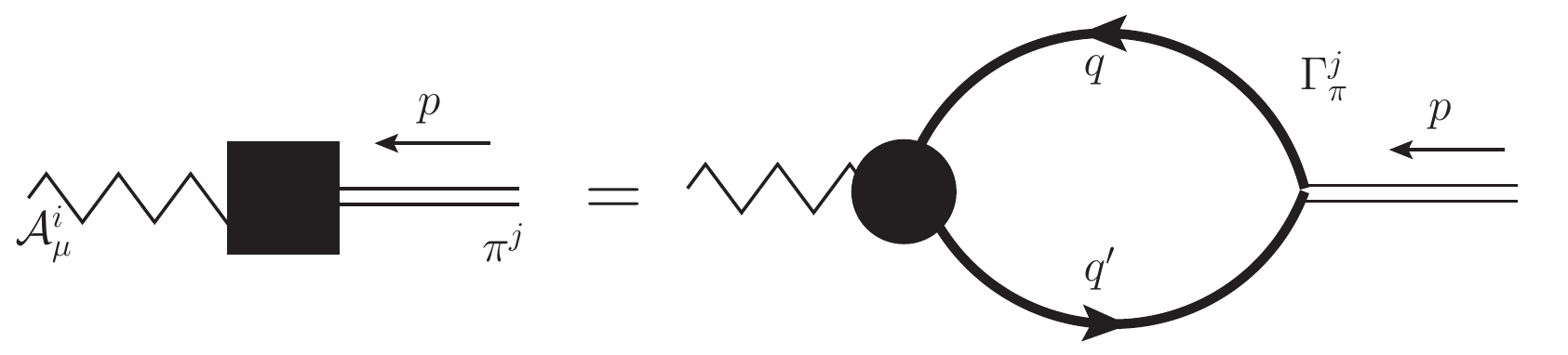}
 \caption{\label{Fig:propAPion2} The ${\cal A}_\mu-\pi$ correlator (black square) in momentum space in terms of the quark propagator and of the axial-vector-quark-antiquark vertex. This is an expression equivalent to the one shown in Fig.~\ref{Fig:propAPion}.}
\end{figure}

Next, consider the vertex Ward identity, derived from Eq.~\eqref{appeq:wardvertex}, relating the pion-quark-antiquark $\pi q\bar q$ and the ${\cal A}_\mu q\bar q$ vertices: 
\begin{align}
\label{appeq:chiralWard}
 &{\cal M}_\Lambda\Gamma_\pi^i(q,q')-\frac{ip_\mu}{2}\Gamma_{{\cal A}_\mu}^i(q,q')\nonumber\\
 &=S^{-1}(q)\frac{i\gamma_5\sigma^i}{2}+\frac{i\gamma_5\sigma^i}{2}S^{-1}(q'),
\end{align}
where $\smash{p=q-q'}$ denotes the incoming pion or axial-vector momentum.\footnote{Our conventions are such that, at tree level, $\Gamma_\pi^i(q,q')\to i\gamma_5\sigma^i$ and $\Gamma_{{\cal A}_\mu}^i(q,q')\to i\gamma_\mu\gamma_5\sigma^i$. Note that isospin symmetry guarantees that the flavor structure of both the pion vertex is $\Gamma_\pi^i(q,q')=i\gamma_5\sigma^i\Gamma_\pi(q,q')$ and similarly for $\Gamma_{{\cal A}_\mu}^i$. Finally, note that the identity \eqref{appeq:Wardpropag} can be obtained from the vertex identity \eqref{appeq:chiralWard} using the exact relations
\begin{align}
 G^{ij}_{\pi\pi}(p)&=-\int_q \text{tr}\left[\Gamma_\pi^i(q',q)S(q)i\gamma_5\sigma^jS(q')\right],\nonumber\\
 G^{ij}_{{\cal A}_\mu\pi}(p)&=-\int_q \text{tr}\left[\Gamma_{{\cal A}_\mu}^i(q',q)S(q)i\gamma_5\sigma^jS(q')\right]\nonumber,
\end{align}
where, by convention, $p$ is the incoming pion momentum in both cases, hence the outgoing axial-vector momentum. Figure \ref{Fig:propAPion2} shows the diagrammatic representation of the second equation above, equivalent to the one shown in Fig.~\ref{Fig:propAPion}.
The relations above express identities such as
\begin{align}
 \frac{\delta^2\Gamma}{\delta J^i(x)\delta J^j(y)}&=-\frac{\delta\langle\pi^j(y)\rangle}{\delta J^i(x)}={\rm tr}\left[i\gamma_5\sigma^j \frac{\delta S_{\cal J}(y,y)}{\delta J^i(x)}\right]\nonumber\\
 &=-\int_{z,z'}{\rm tr}\left[i\gamma_5\sigma^j  S_{\cal J}(y,z)\frac{\delta S^{-1}_{\cal J}(z,z')}{\delta J^i(x)} S_{\cal J}(z',y)\right],
\end{align}
with $S_{\cal J}$ the quark propagator and $\delta S^{-1}_{\cal J}/\delta J^i$ is the pion vertex in the presence of sources. A similar identity involving the the axial-vector current holds.
}
From Eqs.~(\ref{eq:vertexpole}) and (\ref{eq:lorentzdec}), we have $\Gamma_\pi^i(q,q)=i\gamma_5\sigma^i\gamma_P(q^2)/{\cal M}_\Lambda$. Thus, taking the limit $p\to 0$ in Eq.~(\ref{appeq:chiralWard}), and under the assumption that $\Gamma_{{\cal A}_\mu}^i(q,q)$ is regular, this directly yields
\begin{align}
 \gamma_P(q^2)=\frac{M(q^2)}{Z(q^2)}.
\end{align}

Another consequence of the chiral Ward identities is the relation between the rainbow and the ladder integral equations for the quark propagator and the pion or axial-vector vertices, respectively, see Figs.~\ref{Fig:eq_rainbow} and \ref{Fig:BSE}. The former writes
\begin{align}
\label{appeq:RainbowS}
 S^{-1}(q)=-i\slashed{q}+{\cal M}_\Lambda+\lambda_\Lambda\!\int_k  G_{\rho\sigma}(k)\gamma_\rho S(\ell)\gamma_\sigma,
\end{align}
and the latter are\footnote{The relation between rainbows and ladders follows directly from the general relation
\begin{align}
 \Gamma_{\pi}^i(q,q')=-\left.\frac{\delta S_{\cal J}^{-1}(q,q')}{\delta J^i(p)}\right|_{{\cal J}=0}
\end{align}
where $S_{\cal J}^{-1}(q,q')$ is the (nondiagonal) momentum space quark propagator in presence of the source term \eqref{appeq:chiralsourceterm}. The rainbow resummation for the latter reads
\begin{align}
 S^{-1}_{\cal J}(q,q')&={\cal S}^{-1}(q,q')-J^i(p) i\gamma_5\sigma^i\nonumber\\
 &+\lambda_\Lambda\!\int_k  G_{\rho\sigma}(k)\gamma_\rho S_{\cal J}(\ell,\ell')\gamma_\sigma,
\end{align}
with $\ell=q-k$ and $\ell'=q'-k$ and where ${\cal S}^{-1}(q,q')=(-i\slashed{q}+{\cal M}_\Lambda)(2\pi)^4\delta^{(4)}(q-q')$ is the tree-level propagator. Deriving with respect to the source gives and setting it to zero gives Eq.~\eqref{appeq:BSscalar}. A similar treatment leads to Eq.~\eqref{appeq:BSAxial}.
}
\begin{align}
\label{appeq:BSscalar}
\Gamma_{\pi}^i(q,q')&=i\gamma_5\sigma^i-\lambda_\Lambda\!\int_k  G_{\rho\sigma}(k)\gamma_\rho S(\ell)\Gamma_{\pi}^i(\ell,\ell')S(\ell')\gamma_\sigma,\\
\label{appeq:BSAxial}
\Gamma_{{\cal A}_\mu}^i(q,q')&=i\gamma_\mu\gamma_5\sigma^i\nonumber\\
&-\lambda_\Lambda\!\int_k  G_{\rho\sigma}(k)\gamma_\rho S(\ell)\Gamma_{{\cal A}_\mu}^i(\ell,\ell')S(\ell')\gamma_\sigma,
\end{align}
with $\ell=q-k$ and $\ell'=q'-k$. One easily verifies that these satisfy the symmetry identity \eqref{appeq:chiralWard}.

\section{Details of linear-order BSE}
\label{Appsec:BSE}

We present here the derivation of the on-shell Bethe-Salpether equations in the chiral limit, Eqs.~\eqn{Eq:smallPgammaT}--\eqn{Eq:smallPgammaB}.
First, let us consider \Eqn{eq:BSfullvertex} at $p^2=0$. Using the definition \eqn{eq:vertexpole}, with \eqn{eq:lorentzdec} and \eqn{eq:expPT}, we have
\begin{align}
\label{appeq:BSvertexM}
 \gamma_P(r^2)={\cal M}_\Lambda+3\lambda\!\int_s  G(k) \frac{Z^2(s^2)}{s^2+M^2(s^2)}\gamma_P(s^2)\,,
\end{align}
with $k=r-s$, which is identical to the integral equation corresponding to the resummation of rainbow diagrams for the quantity $M(x)/Z(x)$, as demanded by the axial Ward identities for any value of ${\cal M}_\Lambda$; see Sec~\ref{Appsec:Ward}. 

Next, we evaluate \Eqn{eq:BSfullvertex} on the pion mass shell, $p^2=-m_\pi^2$, which gives
\begin{align}
\label{appeq:BSvertex}
 \gamma_\pi(q,q')=\lambda\!\int_k  G_{\mu\nu}(k)\gamma_\mu S(-\ell)\gamma_\pi(\ell,\ell')S(\ell')\gamma_\nu\,.
\end{align}
In the chiral limit, we expand at linear order in $p_\mu$ around $p^2=0$. Using the definitions \eqn{eq:expPT} and \eqn{eq:expA}, the left-hand side reads
\begin{align}
 \gamma_\pi(q,q')&=\gamma_P(r^2)+i\sigma_{\mu\nu}p_\mu r_\nu \gamma_T(r^2)\nonumber\\
 &+i\slashed{p}\gamma_A(r^2)+2i \slashed{r}\frac{p\cdot r}{r^2}[\gamma_A(r^2)-\gamma_B(r^2)]\nonumber\\
 &+{\cal O}(p^2),
\end{align}
whereas, upon writing $\ell=s+p/2$ and $\ell'=s-p/2$ for the integrand on the right-hand side, we obtain, after some algebra,
\begin{align}
& S(-\ell)\gamma_\pi(\ell,\ell')S(\ell')=\frac{Z(s^2)M(s^2)}{s^2+M^2(s^2)}\nonumber\\
&-i\sigma_{\mu\nu}p_\mu s_\nu N(s^2)-i\slashed{p}H(s^2)+2i\slashed{s}p\cdot sL(s^2)+{\cal O}(p^2),
\end{align}
where we used \Eqn{eq:Wardbare} in the first line and where the functions $N$, $H$, and $L$ are defined in Eqs.~\eqn{eq:Nbare}--\eqn{eq:Lbare}. We then project out the scalar, tensor, and vector components of \Eqn{appeq:BSvertex}. As expected, the scalar part reduces to \Eqn{appeq:BSvertexM} in the limit ${\cal M}_\Lambda\to0$. The tensor and scalar component yields
\begin{widetext}
\begin{align}
 (p_\mu r_\nu-p_\nu r_\mu)\gamma_T(r^2)=\lambda\int_sG(k)N(s^2)\left[p_\mu s_\nu-p_\nu s_\mu-2\frac{k\cdot s}{k^2}(p_\mu k_\nu-p_\nu k_\mu)-2\frac{k\cdot p}{k^2}(k_\mu s_\nu-k_\nu s_\mu)\right]
\end{align}
and
\begin{align}
 p_\mu\gamma_A(r^2)+2\frac{p\cdot r}{r^2}r_\mu[\gamma_A(r^2)-\gamma_B(r^2)]=\lambda\int_sG(k)\left\{H(s^2)\left(p_\mu+2\frac{k\cdot p}{k^2}k_\mu\right)-2p\cdot sL(s^2)\left(s_\mu+2\frac{k\cdot s}{k^2}k_\mu\right)\right\}.
\end{align}

To proceed, we exploit the Euclidean Lorentz symmetry and choose, with no loss of generality, $r_\mu=(0,0,0,r)$ and $p_\mu=(0,0,p_3,p_4)$. Accordingly, we write $s_\mu=(\vec s_\perp,s_3,s_4)$ and $k_\mu=(-\vec s_\perp,-s_3,r-s_4)$ and we note that the functions $G(k)=1/(k^2+m^2)$, $N(s^2)$, $H(s^2)$, and $L(s^2)$ are all even in $s_3$. We can, thus, discard explicit odd powers of $s_3$ in the various integrals. Finally, we choose to systematically eliminate any explicit occurence of $s_4$ in favour of $s^2$, $s_3^2$, and $s_\perp^2$. We obtain, after some algebra
\begin{align}
\label{appeq:gammaT}
 \gamma_T(r^2)&=\lambda\int_sG(k)N(s^2)\left(\frac{r^2+s^2}{2r^2}-\frac{\left(r^2-s^2\right)^2}{2r^2k^2}-\frac{2s_3^2}{k^2}\right)\\
\label{appeq:gammaA}
 \gamma_A(r^2)&=\lambda\int_sG(k)\left\{H(s^2)\left(1+\frac{2s_3^2}{k^2}\right)-2L(s^2)s_3^2\left(2-\frac{r^2-s^2}{k^2}\right)\right\}\\
\label{appeq:gammaB}
 \gamma_B(r^2)&=\lambda\int_sG(k)\left\{H(s^2)\frac{4s_3^2+s_\perp^2}{k^2}+L(s^2)\left(3s^2-8s_3^2-2s_\perp^2+\frac{r^2}{k^2}[2s_3^2-s_\perp^2]-\frac{s^2}{k^2}[4s_3^2+s_\perp^2]\right)\right\}
\end{align}
\end{widetext}
Choosing $s$, $s_\perp$, and $s_4$ as independent variables, we can perform the integrals over $s_\perp$ and $s_4$ explicitly, using
\beq
\int_s=\frac{1}{32\pi^3}\int_0^\infty ds^2\int_0^{s^2} ds^2_\perp\int_{-s_B}^{s_B}\frac{ds_4}{s_3}
\eeq
where $s_3=\sqrt{s^2-s_\perp^2-s_4^2}$ and $s_B=\sqrt{s^2-s_\perp^2}$. We introduce the function
\begin{align}
h_{m^2}(r^2,s^2,s_\perp^2)&=\int_{-s_B}^{s_B} ds_4\frac{s_3}{r^2+s^2+m^2-2 rs_4}\nonumber\\
&=\frac{\pi}{2r^2}\left(b-\sqrt{b^2-r^2(s^2-s_\perp^2)}\right),
\end{align}
with $b=(r^2+s^2+m^2)/2$, in term of which, the relevant integrals for our purposes read
\begin{align}
f_{m^2}(r^2,s^2)&=\int_0^{s^2} ds_\perp^2\int_{-s_B}^{s_B}\frac{ds_4}{2s_3} G(k)\nonumber\\
&=-\int_0^{s^2} ds_\perp^2\partial_{s_\perp^2}h_{m^2}(r^2,s^2,s_\perp^2)\nonumber\\
&=h_{m^2}(r^2,s^2,0)
\end{align}
and
\begin{align}
I_{m^2}(r^2,s^2)&=\int_0^{s^2} ds_\perp^2\int_{-s_B}^{s_B}\frac{ds_4}{2s_3} 2s_3^2G(k)\nonumber\\
&=\int_0^{s^2} ds_\perp^2h_{m^2}(r^2,s^2,s_\perp^2),
\end{align}
whose explicit expressions are given in Eqs.~\eqn{eq:fm} and \eqn{eq:Im}. We also note the identity
\begin{align}
 &\int_0^{s^2} ds_\perp^2\int_{-s_B}^{s_B}\frac{ds_4}{2s_3} s_\perp^2G(k)\nonumber\\
 &=-\int_0^{s^2} ds_\perp^2s_\perp^2\partial_{s_\perp^2}h_{m^2}(r^2,s^2,s_\perp^2)\nonumber\\
 &=I_{m^2}(r^2,s^2)
\end{align}
We have, then,
\begin{align}
  \int_s f(s^2)G(k) &= \frac{1}{16\pi^3}\int_0^\infty ds^2f(s^2)f_{m^2}(r^2,s^2)\\
  \int_s f(s^2)G(k)2s_3^2  &= \frac{1}{16\pi^3}\int_0^\infty ds^2f(s^2) I_{m^2}(r^2,s^2)\\
  \int_s  f(s^2)G(k)s_\perp^2 &= \frac{1}{16\pi^3}\int_0^\infty ds^2f(s^2) I_{m^2}(r^2,s^2)
\end{align}
for any function $f(s^2)$. Also, writing $G(k)/k^2=[1/k^2-G(k)]/m^2$, one has
\begin{align}
  \int_s f(s^2)G(k)\frac{1}{k^2} &= -\frac{1}{16\pi^3}\int_0^\infty ds^2f(s^2)\Delta f_{m^2}(r^2,s^2)\\
  \int_s f(s^2)G(k)\frac{2s_3^2}{k^2}  &= -\frac{1}{16\pi^3}\int_0^\infty ds^2f(s^2) \Delta I_{m^2}(r^2,s^2)\\
  \int_s  f(s^2)G(k)\frac{s_\perp^2}{k^2} &= -\frac{1}{16\pi^3}\int_0^\infty ds^2f(s^2) \Delta I_{m^2}(r^2,s^2)
\end{align}
Inserting these in Eqs.~\eqn{appeq:gammaT}--\eqn{appeq:gammaB}, one finally arrives at Eqs.~\eqn{Eq:smallPgammaT}--\eqn{Eq:smallPgammaB}.

\section{Ultraviolet behavior}
\label{Appsec:UV}

To analyze the leading ultraviolet behavior, we first write
\begin{align}
\label{eq:powerlawT}
 \hat\gamma_T(x)&=\frac{y_T(\ln x)}{x^2},\\
\label{eq:powerlawA}
 \hat\gamma_A(x)&=\frac{y_A(\ln x)}{x},\\ 
\label{eq:powerlawB}
 \hat\gamma_B(x)&=\frac{y_B(\ln x)}{x},
 \end{align}
where the functions $y_{T,A,B}(u)$ are expected to be some power laws at large $u=\ln x$. In this section we assume Eqs. \eqref{eq:NUV}--\eqref{eq:LUV} and check their validity {\it a posteriori}. Under this assumption, the equation for $\gamma_T$ decouples from those of $\gamma_A$ and $\gamma_B$. Let us analyze the former first.

First note ({\it e.g.} using the dominant iterated behaviors of $\gamma_{T,A,B}$) that the integral $\int_0^x$ on the left-hand side of Eq.~\eqref{eq:TUV} is dominated by its upper bound: Separate $\int_0^x=\int_0^{x_0}+\int_{x_0}^x$ with $m^2\ll x_0\ll x$ and check that the contribution $\int_0^{x_0}$ is suppressed by powers of $\ln x$ as compared to that of $\int_{x_0}^x$. We can thus neglect the former and replace the various integrands by their UV behaviors in the latter. Also, for the present analysis it is convenient to momentarily introduce a UV cutoff by replacing $\int_x^\infty\to\int_x^{\Lambda^2}$. Introducing $u_0=\ln x_0$ and $u_\Lambda=\ln\Lambda^2$, we get
\begin{align}
\label{eq:yT}
  y_T(u)&=\frac{A_M}{12} u^{\gamma_M-1}+\frac{\gamma_M}{12 u}\int_{u_0}^u dvy_T(v)\nonumber\\
  &+\frac{\gamma_M}{12 u}\int_u^{u_\Lambda} dve^{2(u-v)}y_T(v).
\end{align}
This can be turned into a second order differential equation. Introducing $z_T(u)=uy_T(u)$, $z_s(u)=(A_M/12)u^{\gamma_M}$, and $\alpha=\gamma_M/12$, we have
\begin{align}
  z_T(u)=z_s(u)+\alpha\!\int_{u_0}^u \frac{dv}{v}z_T(v)+\alpha\!\int_u^{u_\Lambda} \frac{dv}{v}e^{2(u-v)}z_T(v),
\end{align}
where we have introduced an ultraviolet cut-off $u_\Lambda$. One easily checks that
\beq
 z_T''-2z_T'+\frac{2\alpha}{u}z_T=z_s''-2z_s'\sim-2\alpha A_Mu^{12\alpha-1},
\eeq
with the boundary condition 
\beq
\label{Appeq:BCzT}
 z_T'(u_\Lambda)=z_s'(u_\Lambda).
\eeq
 The general solution at large $u$ (keeping $1\ll u\ll u_{\Lambda}$) is, for $\alpha>0$,
\beq
 z_T\sim\frac{A_M}{11}u^{12\alpha}+{ A}_Tu^{\alpha}+{ B}_Te^{2u}u^{-\alpha}
\eeq
with ${ A}_T$ and ${ B}_T$ some integration constants. All lead to ultraviolet finite integrals. The term $\propto A_T$ is negligible as compared to the (always present) first term so we can write  
\beq
 z_T\sim\frac{A_M}{11}u^{12\alpha}+{ B}_Te^{2u}u^{-\alpha}
\eeq
The constant $B_T$ is obtained from the boundary condition \eqref{Appeq:BCzT} as
\beq
 B_T=-\frac{\alpha A_M}{22} u_\Lambda^{13\alpha-1}e^{-2u_\Lambda}
\eeq
and thus vanishes in the limit $u_\Lambda\to\infty$. We finally have 
\beq
 z_T\sim\frac{A_M}{11}u^{12\alpha}.
\eeq

A similar analysis can be made for the coupled integral equations in the $A-B$ sector. In this case, the contribution $\int_{0}^{u_0}$ amounts to a constant term---called $\bar c_A$ in the following---that must be taken into account in the equation for $z_A$ and to a term $\propto 1/x=e^{-u}$ that can be safely neglected in the equation for $z_B$. As before we use the dominant UV behavior of the various integrands in the contributions $\int_{u_0}^u$ and $\int_{u_0}^{u_\Lambda}$. Introducing the functions
\begin{align}
 I_1(u)&=\alpha\int_{u_0}^u \frac{dv}{v}z_B(v),\\
 I_2(u)&=\alpha\int_{u_0}^u \frac{dv}{v}e^{v-u}[z_A(v)+z_B(v)],\\
 I_3(u)&=\alpha\int_u^{u_\Lambda}  \frac{dv}{v}e^{u-v}[z_B(v)-z_A(v)],\\
 I_4(u)&=\alpha\int_u^{u_\Lambda}  \frac{dv}{v}e^{2(u-v)}[2z_A(v)-z_B(v)],
\end{align}
we have
\begin{align}
\label{appeq:intA}
 z_A(u)&=\bar c_A+I_1(u)+\frac{1}{3}I_2(u)+3I_3(u)+\frac{5}{3}I_4(u)\\
\label{appeq:intB}
 z_B(u)&=I_2(u)+3I_3(u)+2I_4(u)
\end{align}
One checks that these satisfy the following coupled differential equations
\beq
\label{appeq:diffA}
 z_B+2z_B'=3z_A'
\eeq
and
\begin{equation}
\label{appeq:diffB}
 z_B'''-2z_B''-z_B'+2z_B=\frac{6\alpha}{u}\left(\frac{z_A}{u}-z_A'+z_B\right),
 \end{equation}
together with the condition
\beq
 z_B'(u_\Lambda)=3z_A'(u_\Lambda).
\eeq
Alternatively, Eq. \eqref{appeq:diffB} rewrites
\begin{align}
 &z_B'''-2z_B''-\left(1-\frac{4\alpha}{u}\right)z_B'+2\left(1-\frac{2\alpha}{u}\right)z_B=\frac{6\alpha z_A}{u^2}.
 \end{align}
It is an easy matter to find the solutions in an expansion at large $u$. We obtain, for the dominant terms, 
\begin{align}
\label{appeq:dsolA}
z_A&\sim A_1+ A_2e^u+A_3e^{-u}u^{4\alpha/3}+A_4 e^{2u}u^{-4\alpha/3}\\
\label{appeq:dsolB}
z_B&\sim \frac{3\alpha A_1}{u^2}+ A_2e^u+ 3A_3e^{-u}u^{4\alpha/3}+\frac{6}{5}A_4 e^{2u}u^{-4\alpha/3}.
\end{align} 

Clearly, not all solutions of the above differential equations are solution of the original integral equation. To select the required solution, we plug the expressions \eqref{appeq:dsolA} and Eqs. \eqref{appeq:dsolB} back in integral equations  \eqref{appeq:intA} and Eqs. \eqref{appeq:intB}. A consistent solution requires 
$A_1=\bar c_A$ and $A_2=3\alpha A_4e^{u_\Lambda}/(5u_\Lambda^{4\alpha/3+1})$. Requiring a finite solution in the limit $u_\Lambda\to\infty$ therefore implies $A_4=A_2=0$. The term $\propto A_3$ can be neglected because $A_1=\bar c_A\neq0$ and we arrive at 
\begin{equation}
z_A\sim \bar c_A\quad {\rm and} \quad z_B\sim \frac{3\alpha \bar c_A}{u^2}.
\end{equation} 

Finally, the constant $\bar c_A$ can be determined by inserting these results back in Eq. \eqref{eq:AUV}. One easily checks that  
\begin{align}
\hat\gamma_A(x)&=\frac{\lambda(x)}{32 \pi^2}\int_{0}^\infty dyZ_\psi(y)\frac{y}{x}\left[\hat H(y)-\frac{y}{2}\hat L(y)\right] \nonumber\\&+ {\cal O}\left(\frac{1}{x(\ln x)^2}\right),
\end{align} 
where one recognize the integral in Eq. \eqref{Eq:renFpi}. It follows that
\begin{equation}
 \bar c_A=\frac{C_F f_\pi^2}{8N_c\beta_0}=\frac{4\pi^2\gamma_M}{6N_c}f_\pi^2.
\end{equation}

\section{Ultraviolet tails}
\label{Appsec:UVresults}

\begin{figure}[t!]
 \includegraphics[width=0.5\textwidth]{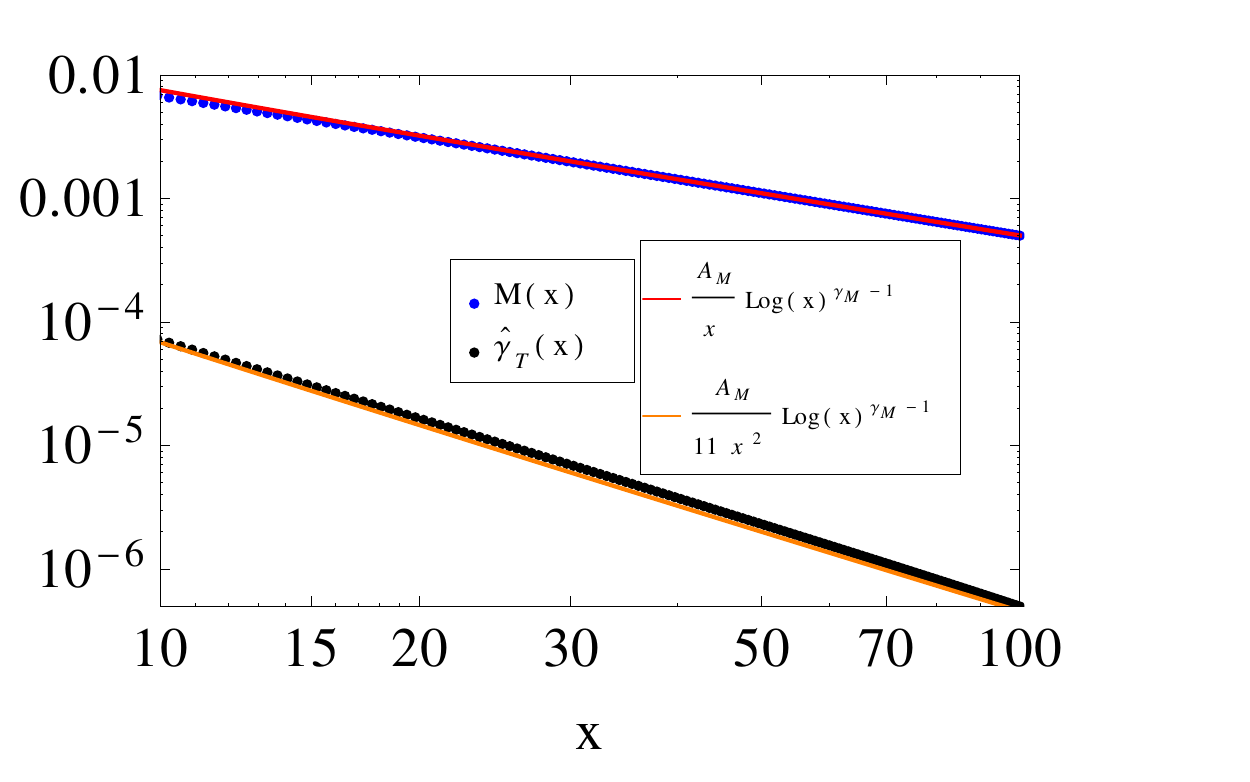}
 \includegraphics[width=0.5\textwidth]{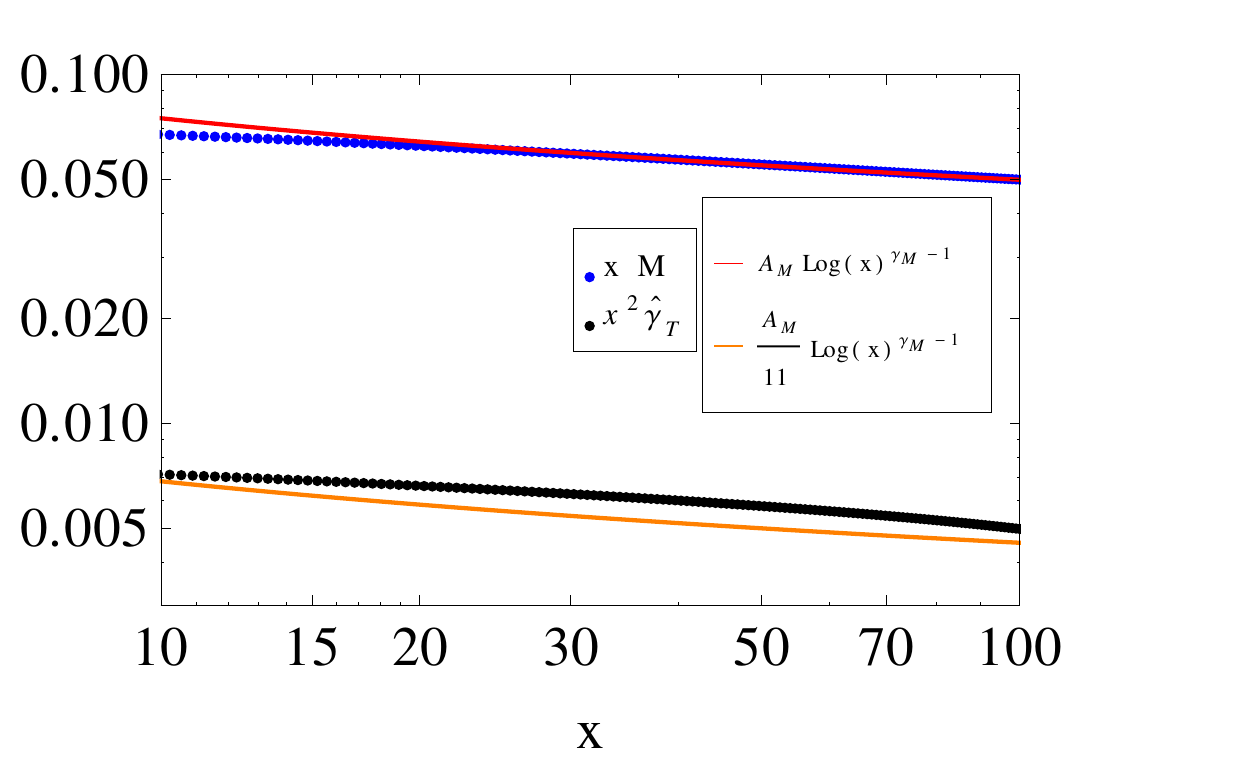}
  \caption{\label{Fig:gammasAjustePT} The large momentum behavior of the functions $M(x)$ and $\hat\gamma_T(x)$ for $g_0=1.93$ and $m_0=0.11$~GeV compared with the expected asymptotic behaviors \eqref{eq:ZMUV} and \eqref{eq:UVgammaT}. We fit the value $A_M=0.12~{\rm GeV}^3$.}
\end{figure}
\begin{figure}[t!]
 \includegraphics[width=0.5\textwidth]{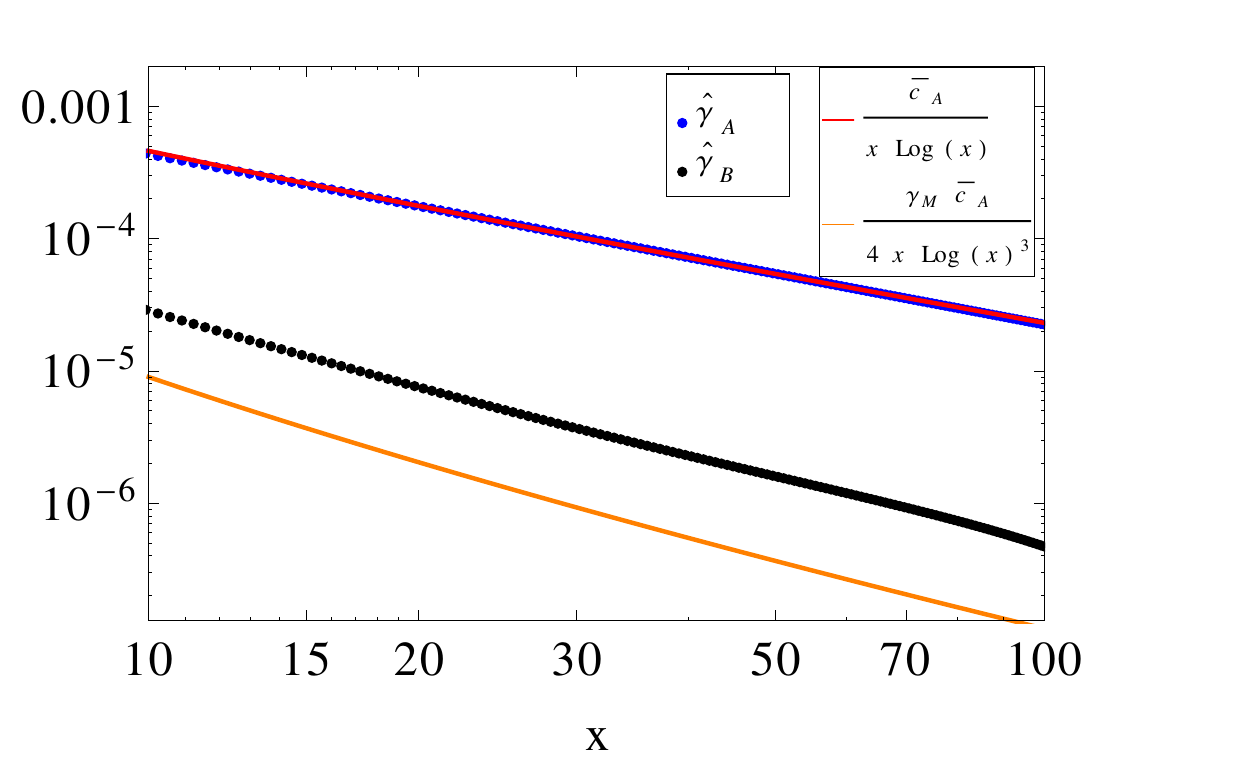}
 \includegraphics[width=0.5\textwidth]{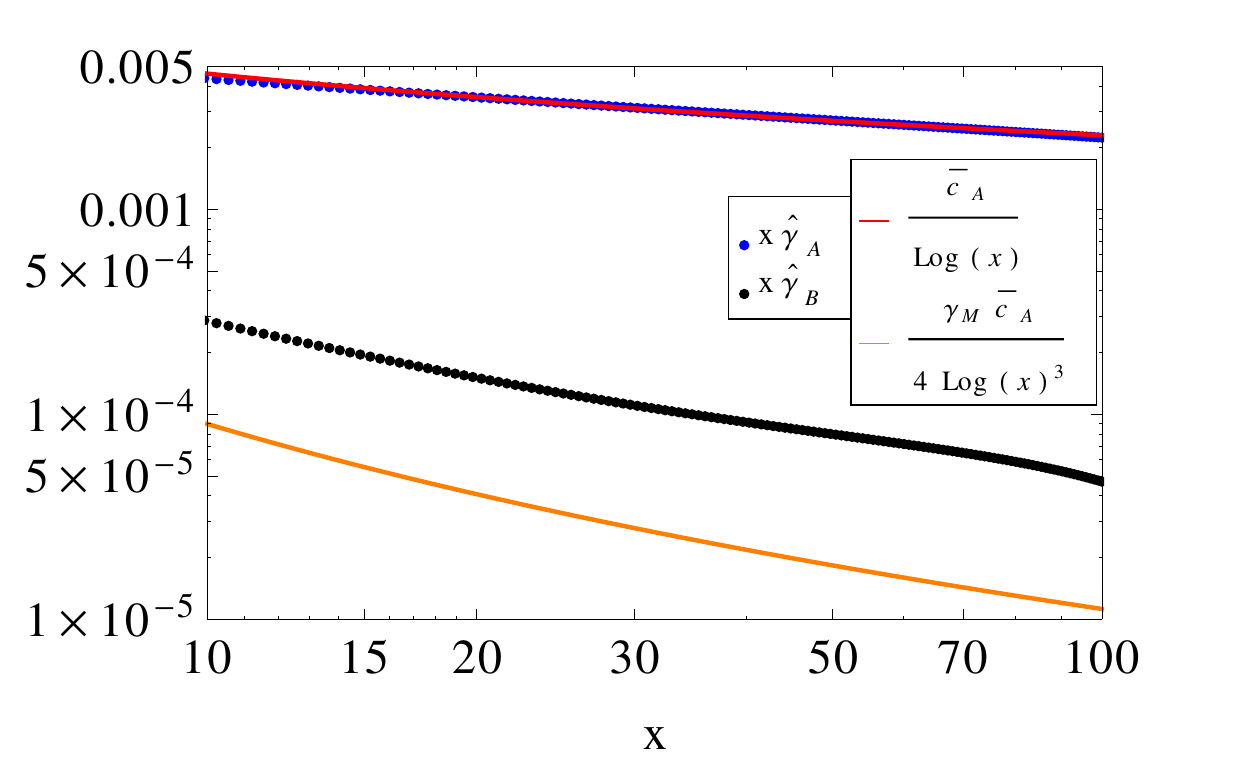}
  \caption{\label{Fig:gammasAjusteAB} The large momentum behavior of the functions  $\hat\gamma_A(x)$ and $\hat\gamma_B(x)$ using $g_0=1.93$ and $m_0=0.11$ GeV compared with the asymptotic behaviors asymptotic behaviors  \eqref{eq:UVgammaA} and \eqref{eq:UVgammaB}. We fit the value $\bar c_A=0.01~{\rm GeV}^2$.}
\end{figure}

We have tested the UV behavior of our numerical results against the analytical results \eqref{eq:ZMUV}, \eqref{eq:UVgammaT}, \eqref{eq:UVgammaA}, and \eqref{eq:UVgammaB}. This is shown in Figs.~\ref{Fig:gammasAjustePT}--\ref{Fig:gammasAjusteAB}. Although we do not have more than essentially a decade in the square-momentum $x$, our result reproduce well the expected power laws and the logarithmic corrections for all the functions but $\hat\gamma_B(x)$. For instance, we observe that the predicted ratio $x\hat\gamma_T/M$ is well reproduced, but not the ratio $\hat\gamma_A/\hat\gamma_B$. We understand this as due to the fact that, as explained in Sec.~\ref{sec:UV}, the behavior \eqref{eq:UVgammaB} arises from a cancelation of the naive leading behavior in Eq.~\eqref{eq:BUV}. Because the UV tails contribution to $f_\pi$ are negligible, see Eq.~\eqref{eq:fpiUV}, we have not attempted to resolve this issue further.

From the UV behaviors of $M$ and $\hat \gamma_A$, we fit the constants $A_M$ and $\bar c_A$, although these should be taken with a grain of salt because those fits are realized over a restricted range of UV momenta. We simply check here that this have the expected orders of magnitude. A detailed analysis would require a dedicated study of the deep UV regime. As recalled in Sec.~\ref{Appsec:condensate}, the constant $A_M$ is related to the renormalized RG-invariant quark condensate in the chiral limit $\tilde\sigma_R$ as ($N_c=3$, $\gamma_M\approx0.4$)
\beq
 A_M=-\frac{2^{1-\gamma_M}\pi^2\gamma_M}{N_c}\tilde\sigma_R\approx-1.99\tilde\sigma_R.
\eeq
For the parameters $g_0=0.193$ and $m_0=0.11~{\rm GeV}$ that give $f_\pi\approx 86~{\rm MeV}$, we fit $A_M\approx 0.12~{\rm GeV}^3$,  $\tilde\sigma_R\approx (392~{\rm MeV})^3$, which is the correct order of magnitude \cite{DeGrand:2006nv,Davies:2012xun,Wang:2016lsv,Aoki:2016frl}. In the parameter space studied here, we find $355~{\rm MeV}\lesssim \tilde\sigma_R^{1/3}\lesssim 411~{\rm MeV}$. 

As for the constant $\bar c_A$, we can compare it to the predicted value 
\beq
 \frac{\bar c_A}{f_\pi^2}=\frac{4\pi^2\gamma_M}{6N_c}\approx 0.87
\eeq
For the same parameters as above, we obtain $\bar c_A\approx 0.01~{\rm GeV}^2$, that is, $\bar c_A/f_\pi^2\approx 1.35$, roughly in the right ballpark.

\section{The quark condensate}
\label{Appsec:condensate}

For completeness, we briefly recall some aspects of the quark condensate in the chiral limit and its relation to the power-law decrease of the quark mass function at large momentum \cite{Fischer:2003rp}. The bare quark condensate $\sigma=\langle\bar\Psi\Psi\rangle$ is UV divergent and requires regularisation. Using a hard cut-off, it reads, in terms of the renormalized quark propagator,
\begin{align}
 \sigma&=-\frac{N_fN_c}{4\pi^2}\int_0^{\Lambda^2}dx\,x\frac{Z_\psi(x)M(x)}{x+M^2(x)}\,.
\end{align}
The integral is controlled by the large-$x$ behavior \eqref{eq:ZMUV} of the integrand and reads
\beq
\label{appeq:sigmabare}
 \sigma=-\frac{N_fN_c}{4\pi^2}\frac{A_M}{\gamma_M}\left(\ln \frac{\Lambda^2}{\Lambda_{\rm QCD}^2}\right)^{\gamma_M},
\eeq
where the scale under the logarithm is arbitrary. 

One defines the renormalized quark condensate as
\beq
\label{appeq:rencond}
 \sigma=Z_\sigma (\mu_0^2)\sigma_R(\mu_0^2)=Z_{\cal M}^{-1} (\mu_0^2)\sigma_R(\mu_0^2)\,,
\eeq
where we used the renormalization condition $\smash{Z_\sigma (\mu_0^2)Z_{\cal M}(\mu_0^2)=1}$, with $Z_{\cal M}$ the quark mass renormalization factor, see the discussion below Eq.~\eqref{eq:Renormgamma}. In the present scheme, the latter is defined as $\smash{{\cal M}_\Lambda=Z_{\cal M}(\mu_0^2)M(\mu_0^2)}$, with ${\cal M}_\Lambda$ the bare quark mass. Although the bare quark mass ${\cal M}_\Lambda$ vanishes in the chiral limit, the renormalization factor $Z_{\cal M}(\mu_0^2)$ has a nontrivial limit, given by the standard RG analysis \cite{Weinberg:1996kr}: At one-loop order, one has, in the UV, $d \ln Z_M/d\ln \mu=2\gamma_M\beta_0g^2(\mu)$, with $g(\mu)$ the running coupling. It follows that $Z_M\propto g^{-2\gamma_M}$ and thus that $\sigma_R g^{2\gamma_M}$ is RG invariant. 

Choosing the renormalization condition $\sigma_R(\Lambda^2)=\sigma$ and defining the RG-invariant condensate 
\beq
 \tilde \sigma_R=\sigma_R(\mu_0)\left[2\beta_0g^2(\mu_0^2)\right]^{\gamma_M}=\frac{\sigma_R(\mu_0)}{\left(\frac{1}{2}\ln\frac{\mu_0^2}{\Lambda_{\rm QCD}^2}\right)^{\gamma_M}},
\eeq
we deduce from Eq.~\eqref{appeq:sigmabare} that
\beq
 A_M=-\frac{2^{1-\gamma_M}\pi^2\gamma_M}{N_c}\tilde\sigma_R.
\eeq
With these definitions, the large-momentum behavior of the quark mass function writes 
\beq
 M(x)\sim\frac{2\pi^2\gamma_M}{N_fN_c}\frac{-\tilde\sigma_R}{x\left(\frac{1}{2}\ln \frac{x}{\Lambda_{\rm QCD}^2}\right)^{1-\gamma_M}}
\eeq
and the running quark condensate is given by
\beq
 \sigma_R(\mu^2)=\tilde \sigma_R\left(\frac{1}{2}\ln\frac{\mu^2}{\Lambda_{\rm QCD}^2}\right)^{\gamma_M}.
\eeq
These reproduce the corresponding expressions in Ref.~\cite{Fischer:2003rp}.

\bibliographystyle{apsrev4-1}
\bibliography{bibli2}

\end{document}